\documentclass[twocolumn,preprintnumbers]{aastex61}
\pdfoutput=1 
\usepackage{amsmath,amstext}
\usepackage[T1]{fontenc}
\usepackage{apjfonts}
\usepackage[figure,figure*]{hypcap}
\usepackage{cleveref}
\usepackage{amsmath,amssymb}
\usepackage{lineno}
\newlength{\colwidth}

\newcommand{\Abreak}{4000 \AA\ }
\newcommand{\hinv}{\ensuremath{h^{-1}}}

\newcommand{\hmsun}{\ensuremath{{h^{-1}}{\rm M}_{\solar}}}
\newcommand{\addgals}{\textsc{Addgals}}
\newcommand{\de}{\text{d}}

\newcommand{\msol}{\ensuremath{{\rm M}_{\solar}}}
\newcommand{\solar}{\ensuremath{_{\mathord\odot}}}
\newcommand{\hmpc}{\ifmmode{h^{-1}{\rm Mpc}}\;\else${h^{-1}}${\rm Mpc}\fi}
\newcommand{\hMpc}{\ifmmode{h^{-1}{\rm Mpc}}\;\else${h^{-1}}${\rm Mpc}\fi}
\newcommand{\hGpc}{\ifmmode{h^{-1}{\rm Gpc}}\;\else${h^{-1}}${\rm Gpc}\fi}
\newcommand{\hkpc}{\ifmmode{h^{-1}{\rm kpc}}\;\else${h^{-1}}${\rm kpc}\fi}
\newcommand{\lcdm}{\ensuremath{\Lambda}CDM}
\newcommand{\msun}{\ensuremath{{\rm M}_{\solar}}}
\newcommand{\LCDM}{\ensuremath{\Lambda}CDM}

\newcommand{\mr}{\ifmmode{M_r}\;\else$M_r$\fi}
\newcommand{\rd}{\ifmmode{R_\delta}\;\else$R_\delta$\fi}
\newcommand{\ngals}{\ifmmode{N_{\rm gals}}\;\else$N_{\rm gals}$\fi}

\newcommand{\comment}[1]{}
\def\vs{\nonumber\\}

\begin{document}
\begin{nolinenumbers}
\vspace*{-\headsep}\vspace*{\headheight}
\footnotesize \hfill FERMILAB-PUB-19-004-AE\\
\vspace*{-\headsep}\vspace*{\headheight}
\footnotesize \hfill DES-2018-0406
\end{nolinenumbers}
\title{The Buzzard Flock: Dark Energy Survey Synthetic Sky Catalogs}
\shorttitle{The Buzzard Flock}
\author{J.~DeRose}
\affiliation{Department of Physics, Stanford University, 382 Via Pueblo Mall, Stanford, CA 94305, USA}
\affiliation{Kavli Institute for Particle Astrophysics \& Cosmology, P. O. Box 2450, Stanford University, Stanford, CA 94305, USA}
\affiliation{SLAC National Accelerator Laboratory, Menlo Park, CA 94025, USA}
\author{R.~H.~Wechsler}
\affiliation{Department of Physics, Stanford University, 382 Via Pueblo Mall, Stanford, CA 94305, USA}
\affiliation{Kavli Institute for Particle Astrophysics \& Cosmology, P. O. Box 2450, Stanford University, Stanford, CA 94305, USA}
\affiliation{SLAC National Accelerator Laboratory, Menlo Park, CA 94025, USA}
\author{M.~R.~Becker}
\affiliation{Argonne National Laboratory, 9700 South Cass Avenue, Lemont, IL 60439, USA}
\author{M.~T.~Busha}
\affiliation{Securiti, mbusha@gmail.com}
\author{E.~S.~Rykoff}
\affiliation{Kavli Institute for Particle Astrophysics \& Cosmology, P. O. Box 2450, Stanford University, Stanford, CA 94305, USA}
\affiliation{SLAC National Accelerator Laboratory, Menlo Park, CA 94025, USA}
\author{N.~MacCrann}
\affiliation{Center for Cosmology and Astro-Particle Physics, The Ohio State University, Columbus, OH 43210, USA}
\affiliation{Department of Physics, The Ohio State University, Columbus, OH 43210, USA}
\author{B.~Erickson}
\affiliation{Department of Physics, University of Michigan, Ann Arbor, MI 48109, USA}
\author{A.~E.~Evrard}
\affiliation{Department of Astronomy, University of Michigan, Ann Arbor, MI 48109, USA}
\affiliation{Department of Physics, University of Michigan, Ann Arbor, MI 48109, USA}
\author{A.~Kravtsov}
\affiliation{Kavli Institute for Cosmological Physics, University of Chicago, Chicago, IL 60637, USA}
\author{D.~Gruen}
\affiliation{Department of Physics, Stanford University, 382 Via Pueblo Mall, Stanford, CA 94305, USA}
\affiliation{Kavli Institute for Particle Astrophysics \& Cosmology, P. O. Box 2450, Stanford University, Stanford, CA 94305, USA}
\affiliation{SLAC National Accelerator Laboratory, Menlo Park, CA 94025, USA}
\author{S.~Allam}
\affiliation{Fermi National Accelerator Laboratory, P. O. Box 500, Batavia, IL 60510, USA}
\author{S.~Avila}
\affiliation{Institute of Cosmology and Gravitation, University of Portsmouth, Portsmouth, PO1 3FX, UK}
\author{S.~L.~Bridle}
\affiliation{Jodrell Bank Center for Astrophysics, School of Physics and Astronomy, University of Manchester, Oxford Road, Manchester, M13 9PL, UK}
\author{D.~Brooks}
\affiliation{Department of Physics \& Astronomy, University College London, Gower Street, London, WC1E 6BT, UK}
\author{E.~Buckley-Geer}
\affiliation{Fermi National Accelerator Laboratory, P. O. Box 500, Batavia, IL 60510, USA}
\author{A.~Carnero~Rosell}
\affiliation{Centro de Investigaciones Energ\'eticas, Medioambientales y Tecnol\'ogicas (CIEMAT), Madrid, Spain}
\affiliation{Laborat\'orio Interinstitucional de e-Astronomia - LIneA, Rua Gal. Jos\'e Cristino 77, Rio de Janeiro, RJ - 20921-400, Brazil}
\author{M.~Carrasco~Kind}
\affiliation{Department of Astronomy, University of Illinois at Urbana-Champaign, 1002 W. Green Street, Urbana, IL 61801, USA}
\affiliation{National Center for Supercomputing Applications, 1205 West Clark St., Urbana, IL 61801, USA}
\author{J.~Carretero}
\affiliation{Institut de F\'{\i}sica d'Altes Energies (IFAE), The Barcelona Institute of Science and Technology, Campus UAB, 08193 Bellaterra (Barcelona) Spain}
\author{F.~J.~Castander}
\affiliation{Institut d'Estudis Espacials de Catalunya (IEEC), 08034 Barcelona, Spain}
\affiliation{Institute of Space Sciences (ICE, CSIC),  Campus UAB, Carrer de Can Magrans, s/n,  08193 Barcelona, Spain}
\author{R.~Cawthon}
\affiliation{Physics Department, 2320 Chamberlin Hall, University of Wisconsin-Madison, 1150 University Avenue Madison, WI  53706-1390}
\author{M.~Crocce}
\affiliation{Institut d'Estudis Espacials de Catalunya (IEEC), 08034 Barcelona, Spain}
\affiliation{Institute of Space Sciences (ICE, CSIC),  Campus UAB, Carrer de Can Magrans, s/n,  08193 Barcelona, Spain}
\author{L.~N.~da Costa}
\affiliation{Laborat\'orio Interinstitucional de e-Astronomia - LIneA, Rua Gal. Jos\'e Cristino 77, Rio de Janeiro, RJ - 20921-400, Brazil}
\affiliation{Observat\'orio Nacional, Rua Gal. Jos\'e Cristino 77, Rio de Janeiro, RJ - 20921-400, Brazil}
\author{C.~Davis}
\affiliation{Kavli Institute for Particle Astrophysics \& Cosmology, P. O. Box 2450, Stanford University, Stanford, CA 94305, USA}
\author{J.~De~Vicente}
\affiliation{Centro de Investigaciones Energ\'eticas, Medioambientales y Tecnol\'ogicas (CIEMAT), Madrid, Spain}
\author{J.~P.~Dietrich}
\affiliation{Excellence Cluster Universe, Boltzmannstr.\ 2, 85748 Garching, Germany}
\affiliation{Faculty of Physics, Ludwig-Maximilians-Universit\"at, Scheinerstr. 1, 81679 Munich, Germany}
\author{P.~Doel}
\affiliation{Department of Physics \& Astronomy, University College London, Gower Street, London, WC1E 6BT, UK}
\author{A.~Drlica-Wagner}
\affiliation{Fermi National Accelerator Laboratory, P. O. Box 500, Batavia, IL 60510, USA}
\affiliation{Kavli Institute for Cosmological Physics, University of Chicago, Chicago, IL 60637, USA}
\author{P.~Fosalba}
\affiliation{Institut d'Estudis Espacials de Catalunya (IEEC), 08034 Barcelona, Spain}
\affiliation{Institute of Space Sciences (ICE, CSIC),  Campus UAB, Carrer de Can Magrans, s/n,  08193 Barcelona, Spain}
\author{J.~Frieman}
\affiliation{Fermi National Accelerator Laboratory, P. O. Box 500, Batavia, IL 60510, USA}
\affiliation{Kavli Institute for Cosmological Physics, University of Chicago, Chicago, IL 60637, USA}
\author{J.~Garc\'ia-Bellido}
\affiliation{Instituto de Fisica Teorica UAM/CSIC, Universidad Autonoma de Madrid, 28049 Madrid, Spain}
\author{G.~Gutierrez}
\affiliation{Fermi National Accelerator Laboratory, P. O. Box 500, Batavia, IL 60510, USA}
\author{W.~G.~Hartley}
\affiliation{Department of Physics \& Astronomy, University College London, Gower Street, London, WC1E 6BT, UK}
\affiliation{Department of Physics, ETH Zurich, Wolfgang-Pauli-Strasse 16, CH-8093 Zurich, Switzerland}
\author{D.~L.~Hollowood}
\affiliation{Santa Cruz Institute for Particle Physics, Santa Cruz, CA 95064, USA}
\author{B.~Hoyle}
\affiliation{Max Planck Institute for Extraterrestrial Physics, Giessenbachstrasse, 85748 Garching, Germany}
\affiliation{Universit\"ats-Sternwarte, Fakult\"at f\"ur Physik, Ludwig-Maximilians Universit\"at M\"unchen, Scheinerstr. 1, 81679 M\"unchen, Germany}
\author{D.~J.~James}
\affiliation{Harvard-Smithsonian Center for Astrophysics, Cambridge, MA 02138, USA}
\author{E.~Krause}
\affiliation{Department of Astronomy/Steward Observatory, 933 North Cherry Avenue, Tucson, AZ 85721-0065, USA}
\author{K.~Kuehn}
\affiliation{Australian Astronomical Optics, Macquarie University, North Ryde, NSW 2113, Australia}
\author{N.~Kuropatkin}
\affiliation{Fermi National Accelerator Laboratory, P. O. Box 500, Batavia, IL 60510, USA}
\author{M.~Lima}
\affiliation{Departamento de F\'isica Matem\'atica, Instituto de F\'isica, Universidade de S\~ao Paulo, CP 66318, S\~ao Paulo, SP, 05314-970, Brazil}
\affiliation{Laborat\'orio Interinstitucional de e-Astronomia - LIneA, Rua Gal. Jos\'e Cristino 77, Rio de Janeiro, RJ - 20921-400, Brazil}
\author{M.~A.~G.~Maia}
\affiliation{Laborat\'orio Interinstitucional de e-Astronomia - LIneA, Rua Gal. Jos\'e Cristino 77, Rio de Janeiro, RJ - 20921-400, Brazil}
\affiliation{Observat\'orio Nacional, Rua Gal. Jos\'e Cristino 77, Rio de Janeiro, RJ - 20921-400, Brazil}
\author{F.~Menanteau}
\affiliation{Department of Astronomy, University of Illinois at Urbana-Champaign, 1002 W. Green Street, Urbana, IL 61801, USA}
\affiliation{National Center for Supercomputing Applications, 1205 West Clark St., Urbana, IL 61801, USA}
\author{C.~J.~Miller}
\affiliation{Department of Astronomy, University of Michigan, Ann Arbor, MI 48109, USA}
\affiliation{Department of Physics, University of Michigan, Ann Arbor, MI 48109, USA}
\author{R.~Miquel}
\affiliation{Instituci\'o Catalana de Recerca i Estudis Avan\c{c}ats, E-08010 Barcelona, Spain}
\affiliation{Institut de F\'{\i}sica d'Altes Energies (IFAE), The Barcelona Institute of Science and Technology, Campus UAB, 08193 Bellaterra (Barcelona) Spain}
\author{R.~L.~C.~Ogando}
\affiliation{Laborat\'orio Interinstitucional de e-Astronomia - LIneA, Rua Gal. Jos\'e Cristino 77, Rio de Janeiro, RJ - 20921-400, Brazil}
\affiliation{Observat\'orio Nacional, Rua Gal. Jos\'e Cristino 77, Rio de Janeiro, RJ - 20921-400, Brazil}
\author{A.~A.~Plazas}
\affiliation{Jet Propulsion Laboratory, California Institute of Technology, 4800 Oak Grove Dr., Pasadena, CA 91109, USA}
\author{A.~K.~Romer}
\affiliation{Department of Physics and Astronomy, Pevensey Building, University of Sussex, Brighton, BN1 9QH, UK}
\author{E.~Sanchez}
\affiliation{Centro de Investigaciones Energ\'eticas, Medioambientales y Tecnol\'ogicas (CIEMAT), Madrid, Spain}
\author{R.~Schindler}
\affiliation{SLAC National Accelerator Laboratory, Menlo Park, CA 94025, USA}
\author{S.~Serrano}
\affiliation{Institut d'Estudis Espacials de Catalunya (IEEC), 08034 Barcelona, Spain}
\affiliation{Institute of Space Sciences (ICE, CSIC),  Campus UAB, Carrer de Can Magrans, s/n,  08193 Barcelona, Spain}
\author{I.~Sevilla-Noarbe}
\affiliation{Centro de Investigaciones Energ\'eticas, Medioambientales y Tecnol\'ogicas (CIEMAT), Madrid, Spain}
\author{M.~Smith}
\affiliation{School of Physics and Astronomy, University of Southampton,  Southampton, SO17 1BJ, UK}
\author{E.~Suchyta}
\affiliation{Computer Science and Mathematics Division, Oak Ridge National Laboratory, Oak Ridge, TN 37831}
\author{M.~E.~C.~Swanson}
\affiliation{National Center for Supercomputing Applications, 1205 West Clark St., Urbana, IL 61801, USA}
\author{G.~Tarle}
\affiliation{Department of Physics, University of Michigan, Ann Arbor, MI 48109, USA}
\author{V.~Vikram}
\affiliation{Argonne National Laboratory, 9700 South Cass Avenue, Lemont, IL 60439, USA}
\collaboration{(DES Collaboration)}

\begin{abstract}
We present a suite of 18 synthetic sky catalogs designed to support science analysis of galaxies in the Dark Energy Survey Year 1 (DES Y1) data. For each catalog, we use a computationally efficient empirical approach, \textsc{Addgals}, to embed galaxies within light-cone outputs of three dark matter simulations that resolve halos with masses above $\sim 5 \times 10^{12} \hinv \msol$ at $z \le 0.32$ and $10^{13} \hinv \msol$ at $z\sim 2$.  The embedding method is tuned to match the observed evolution of galaxy counts at different luminosities as well as the spatial clustering of the galaxy population. Galaxies are lensed by matter along the line of sight --- including magnification, shear, and multiple images --- using \textsc{calclens}, an algorithm that calculates shear with 0.42 arcmin resolution at galaxy positions in the full catalog. The catalogs presented here, each with the same \LCDM\ cosmology (denoted Buzzard), contain on average $820$ million galaxies over an area of 1120 square degrees with positions, magnitudes, shapes, photometric errors, and photometric redshift estimates. We show that the weak-lensing shear catalog, \textsc{redMaGiC} galaxy catalogs and \textsc{redMaPPer} cluster catalogs provide plausible realizations of the same catalogs in the DES Y1 data by comparing their magnitude, color and redshift distributions, angular clustering, and mass-observable relations, making them useful for testing analyses that use these samples. We make public the galaxy samples appropriate for the DES Y1 data, as well as the data vectors used for cosmology analyses on these simulations.

\end{abstract}

\section{Introduction}

The laboratory provided to us by the night sky has enabled great advances in our understanding of the universe and the laws that govern it. In particular,
astronomical observations currently provide the only evidence for the existence
of dark matter and dark energy \citep{zwicky1933, rubin1980, planck2018,Perlmutter99,Riess98}, and may provide one of the only avenues for studying energies near the Planck scale in the foreseeable future by measuring observable signals related to cosmic inflation \citep[e.g.][]{Arkani-Hamed2015, Abazajian2016}. In the near future, ongoing and next generation galaxy surveys will measure tens of millions of spectra and image tens of billions of galaxies in order to precisely constrain the properties of dark matter, dark energy, neutrinos and inflation. These surveys include the Dark Energy Survey (DES)\footnote{\url{https://www.darkenergysurvey.org}}, Kilo-Degree Survey (KiDS)\footnote{\url{http://kids.strw.leidenuniv.nl}}, Hyper Suprime Cam (HSC) \footnote{\url{http://www.subarutelescope.org/Projects/HSC}}, Dark Energy Spectroscopic Instrument (DESI) \footnote{\url{https://www.desi.lbl.gov/}}, Prime Focus Spectrograph (PFS) \footnote{\url{https://pfs.ipmu.jp/}}, Large Synoptic Survey Telescope (LSST)\footnote{\url{http://www.lsst.org}}, Euclid\footnote{\url{http://http://sci.esa.int/euclid/}} and the Wide Field InfraRed Survey Telescope (WFIRST)\footnote{\url{http://wfirst.gsfc.nasa.gov}}.

Succeeding in this endeavor will require systematics associated with cosmological observables, astrophysical, theoretical and observational, to be controlled at an exquisite level. Further, with many cosmological probes computed from the same data, the characterization of common sources of systematic error is a crucial priority.

A primary avenue for understanding systematic errors will be through the
analysis of synthetic or ``mock'' catalogs. These catalogs attempt to simulate, at varying levels of fidelity, the full range of physical processes that influence various observables in large-area sky surveys, including galaxy fluxes, sizes, and shapes. Such synthetic catalogs are not designed to calibrate the tools used in actual survey analysis. Rather, they provide a development
environment that supports quantitative investigation of sources of systematic
error in specific, model-dependent scenarios. Fundamentally, these catalogs are
\emph{plausible} rather than \emph{definitive} expectations for a given
cosmology.

Ideally, these synthetic catalogs would be constructed using methods able to
predict the intrinsic distribution and properties of galaxies via {\sl ab
initio} solutions of coupled dark matter and baryonic evolution \citep[e.g.][]{Evrard88}.
Progress is being made on this front, both in running large lower resolution
simulations \citep{Schaye2015, Springel2018, Naiman2018, Marinacci2018, Nelson2018, Pillepich2018}, and more detailed high resolution zoom-in simulations 
\citep{Kim2014, Hopkins2018}, but it will be many years before these approaches
can simulate the volumes being observed by wide-field galaxy surveys at the 
necessary resolution while reproducing observed galaxy populations with as good
of fidelity as empirical models.

In the interim, a more practical alternative that has had success is
to place modeled galaxies in their associated dark matter structures using
an empirical or phenomenological model such as subhalo abundance matching related methods \citep{tasitsiomi_etal:04,conroy_etal:06,hearin2013,Crocce2015,moster2017,behroozi2018} or halo occupation distribution (HOD) models \citep{Jing1998, Seljak2000, Yang2003, Berlind2002, Zheng2005, Mandelbaum2006, vandenBosch2007, Zehavi2011, Carretero2015, Zu2015}. The best of these empirical approaches are able to precisely reproduce luminosity functions, star formation histories, and observed luminosity and color dependent clustering of galaxies, something which no other model of galaxy formation can achieve \citep{moster2017, behroozi2018}. Semi analytic models (SAMs) \citep{White1991, Kauffmann1993, Somerville1999, Cole2000, Bower2006, Guo2013, Benson02} attempt to include more physics than the empirical models mentioned above and so in principle are more predictive, but in doing so they have struggled to reproduce some of the observables to the same level of fidelity as empirical models. These various approaches to modeling the galaxy--halo connection have recently been reviewed by \citet{Wechsler2018}.

\begin{figure*}
  \includegraphics[width=\textwidth]{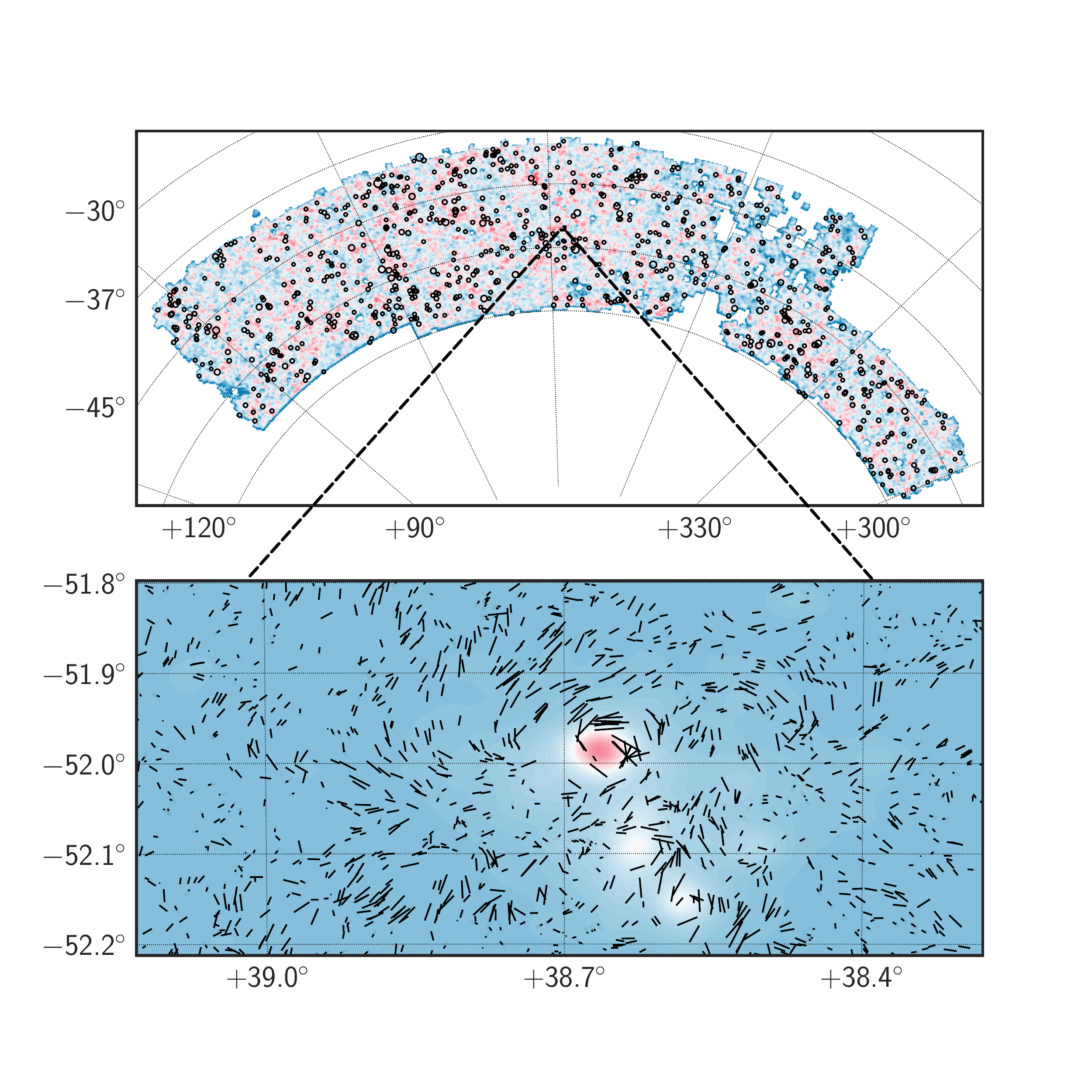}
  \caption{A synthetic Dark Energy Survey sky. \textit{Top}: Projected matter density
  field of one Buzzard footprint, corresponding to the SPT area of the DES Y1
  footprint, with $\lambda>50$ redMaPPer clusters plotted as the black circles.
  Many analyses have used only the portion of this footprint with
  $\textrm{RA}<300$. \textit{Bottom}: Zoom in of a massive halo. The color map represents 
  matter density and black whiskers are the direction and amplitude of the true
  shear of background galaxies in our simulated Y1 weak-lensing catalogs.}
  \label{fig:buzzard_y1_map}
\end{figure*}

In this work, we present a suite of synthetic catalogs for the Dark Energy
Survey (DES) constructed using a model tuned to match the properties of 
sub-halo abundance matching (SHAM), and compare our results to the first year
of DES data (DES Y1). Our methodology, while currently tuned to the DES, is 
applicable generically to a variety of large-area photometric and spectroscopic surveys.
Instead of employing a single, large
N-body simulation, we take a lightweight approach involving sets of smaller dark
matter-only simulations that are readily generated on teraflop computing
platforms. We then apply the \addgals\ algorithm, described in detail in the companion to this work, \citet{wechsler_etal:19}, to populate these dark matter simulations with galaxies. Post-processing routines, including ray-tracing to compute weak-lensing, add a number of physical and instrumental effects, which cause properties of the observed galaxy population to deviate from their intrinsic values. Fig.~\ref{fig:buzzard_y1_map} shows an example of one of our simulations, displaying the underlying projected dark matter distribution, an observed optical cluster catalog, and weak-lensing shear around a massive halo.

In principle, \addgals\ can be used on even lower resolution simulations than the ones presented here, including those produced by approximate $N$-body methods such as COLA \citep{Tassev2013, Izard2018}. The main requirement on the resolution of the simulations used by \addgals\ is that halos above $\sim 10^{13}\hmsun$ be resolved well, driven by the need to resolve the halos hosting galaxy cluster populations. Modern approximate $N$-body codes are capable of reproducing the correct number densities and large-scale clustering of these objects, but the small-scale density profiles of halos suffer from resolution effects. As such, the use case for approximate $N$-body simulations is restricted to applications which do not sensitively depend on small scale galaxy or matter density profiles. As one of the goals of the simulations presented here is to reproduce all of the galaxy-based cosmological observables in DES, including the abundances of optically selected clusters as a function of cluster richness, a quantity that depends sensitively on galaxy profiles in halos, we have forgone using approximate $N$-body simulations, opting instead for traditional $N$-body simulations with modest resolution.

The work presented here is most similar in approach to the MICE simulation \citep{Fosalba2015a, Crocce2015, Fosalba2015b}, with the main qualitative differences being that our methodology allows for the use of lower resolution simulations, and that we include full ray-tracing rather than using the Born approximation to compute weak-lensing observables. Simulations more focused on weak-lensing statistics using full ray-tracing have also recently been released \citep{takahashi2017, harnois-deraps2018}. 

The catalogs we present here are particularly useful because they can be used to
study many large-scale structure probes simultaneously, including galaxy
clustering, optically selected galaxy clusters, weak--lensing shear correlation functions, and the lensing profiles of galaxies and clusters. Further, the total computing time for both the numerical simulations and the post-processing steps is approximately 150K CPU hours per 10,313 square degrees of unique, contiguous sky. This modest scale has allowed us to produce multiple such sky surveys already, and will allow for the production of many more in the near future to meet the needs of DES analyses. Multiple realizations are essential for testing the statistical performance of cosmological analyses and studying the covariances of cosmological observables. 

Indeed, earlier versions of the catalogs presented here have already been essential
for a variety of purposes: development of the DES data management pipeline, testing
and improvement of galaxy cluster finders,
\citep{miller_etal:05,koester_etal:07, dong_etal:08, Hao10,
soares-santos_etal:11}, development of methods to measure cosmological
parameters from galaxy clusters \citep{koester_etal:07b, johnston_etal:07a,
rozo_etal:07, rozo_etal:07b, becker_etal:07, sheldon_etal:09b, hansen_etal:09,
tinker_etal:11}, development and testing of photometric redshift algorithms
\citep{Gerdes10, Cunha12,  Cunha12b, Bonnett15, Leistedt15, Hoyle17, gatti17},
development of various approaches using galaxy shear \citep{Becker2016, Troxel2018, Chang2018}, and in testing the robustness of DES Y1 cosmology pipelines
\citep{y1kp, Krause17, MacCrann17, Friedrich17, Gruen17}. 

This work serves as an explication of the general methodology behind
the production of these simulations, pieces of which have been progressively
improved over the past decade. While the methodology behind these simulations
is still under active development, the versions of the simulations presented
here represent the current state of our modeling capabilities.
In order to assure the usability of simulations for various analyses,
it is vital that working groups familiar with the needs of individual 
analyses contribute quality assurance (QA) requirements that simulations 
must meet. Such an exercise was pursued on a qualitative basis in the
construction of the catalogs presented here, and so the comparisons to
data presented in this work will not be accompanied by quantitative
pass/fail verdicts. Instead we will emphasize where these catalogs have 
found most use within DES, and caution the reader about aspects that 
are particularly untrustworthy. A much more rigorous QA exercise 
is being pursued within the LSST Dark Energy Science Collaboration
(LSST DESC), and the authors are actively engaged in that work
\citep{Mao2018}.

We begin in \S\ref{sec:simreqs} by describing the observables that we wish to simulate. In \S\ref{sec:algorithm} we give a brief summary of the algorithms we applied to produce each synthetic catalog. Note that for clarity, we present the majority of the technical details in the appendices, and in a companion paper describing the implementation and performance of \addgals\ in detail \citep{wechsler_etal:19}. In \S\ref{sec:valid}, we describe our simulated weak-lensing source and lens samples and a photometrically selected cluster sample, comparing them to their analogs in the DES Y1 data. In \S\ref{sec:summary} we summarize and discuss the areas that are most in need of further investigation. Throughout this manuscript, we quote magnitudes using the AB system and $h = 1.0$ units.

\section{Simulation Requirements}
\label{sec:simreqs}

The simulations presented in this work are intended to model observables that 
are used in a range of DES analyses, but, in practice, the analyses that place
the most stringent requirements on simulation fidelity are the key cosmology analyses, 
including cosmic shear, galaxy--galaxy lensing, angular clustering and the combination thereof, called a $3\times 2$-point analysis. Measurements of galaxy cluster abundances will also be a powerful cosmological probe in DES \citep{Weinberg2013,Costanzi2018}. For this reason, we will focus on comparisons between our simulations and the most important quantities for these analyses in this work. 
In particular we will focus on four main galaxy samples in this paper:
\begin{itemize}
\item A sample of galaxies with photometric errors and photometric redshifts approximating the \textsc{Gold} sample.
\item The \textsc{redMaGiC} sample, which was used as a lens galaxy sample in \citet{y1kp}.
\item A weak-lensing source sample, approximating the \textsc{metacalibration} sample used in \citet{y1kp}.
\item The \textsc{redMaPPer} photometric cluster sample.
\end{itemize}
The \textsc{Gold} catalog is the parent catalog of reliable galaxy detections from which all other galaxy samples are selected for DES analyses \citep{y1gold}. The \textsc{redMaGiC} \citep{Rozo2015} and \textsc{redMaPPer} catalogs \citep{rykoff_etal:13} are photometrically selected luminous red galaxy and cluster samples. The \textsc{redMaGiC} sample has been optimized to provide accurate red--sequence--based photometric redshifts and a constant comoving number density.  Likewise, the \textsc{redMaPPer} cluster sample has been optimized to have accurate photometric redshifts, and the richness, $\lambda$, associated with every cluster is a low-scatter halo mass proxy \citep{McClintock2018}. The \textsc{metacalibration} sample in the DES Y1 data is a subsample of the \textsc{gold} catalog with robust ellipticity measurements, made using the \textsc{metacalibration} shear measurement algorithm \citep{Sheldon17,Huff17}, that can be used in the measurement of weak-lensing statistics \citep{Zuntz2017}.

When showing measurements for the samples listed above, we will typically use the redshift bins from \citet{y1kp}, i.e. five lens bins and four source bins for \textsc{redMaGic} and our weak-lensing source sample respectively. In this section we define a number of observables that we wish to model and discuss the process by which they are included in our simulations in the next section.

\subsection{Galaxy Clustering}
As a photometric survey, the primary clustering statistic used in DES is angular
clustering. In particular we are interested in auto- and cross correlations of a galaxy sample binned tomographically by redshift. Assuming the Limber approximation \citep{limber1954}, which is appropriate for the broad redshift binning used for DES observables, and following the notation in \citet{Krause17}, the angular clustering signal is given by:

\begin{align}
w^{i,j}(\theta) &=& \int \frac{dl\,l}{2\pi} J_0(l\theta)\int d\chi \, \frac{q_{\delta_{\mathrm{g}}}^i\!\!\left(\!\frac{l+1/2}{\chi},\chi\right)q_{\delta_{\mathrm{g}}}^j \!\left(\frac{\
l+1/2}{\chi},\chi\right)}{\chi^2}\vs
&&\times P_{\mathrm{NL}}\!\!\left(\frac{l+1/2}{\chi},z(\chi)\!\!\right),
\end{align}
where the radial weight function for clustering in terms of the comoving radial distance $\chi$ is
\begin{equation}
q_{\delta_{\mathrm{g}}}^i(k,\chi) = b^i\left(k,z(\chi)\right)\frac{n_{\mathrm{g}}^i(z(\chi)) }{\bar{n}_{\mathrm{g}}^i}\frac{dz}{d\chi}\,,
\end{equation}
where $J_0$ is the $0$th order Bessel function, redshift distributions given by $n_{\mathrm{g}}^i(z)$, $b^i(k,z(\chi))$ denoting the bias of the galaxies in tomographic bin $i$, $P_{\mathrm{NL}}(k,z)$ the non-linear matter power spectrum at wave number $k$ and redshift $z$, and average angular number densities given by:

\begin{equation}
\bar{n}_{\mathrm{g}}^i = \int dz\; n_{\mathrm{g}}^i(z) \,.
\end{equation}
A number of important details which must be modeled in our simulations become apparent,
including non-linear evolution of the matter distribution, scale-dependent galaxy bias,
and galaxy redshift distributions. 

\subsection{Galaxy Lensing}
We also wish to model the weak-lensing statistics most commonly used in DES. 
In this paper we will discuss measurements of cosmic shear and galaxy--galaxy
lensing. Cosmic shear auto- and cross-correlation functions can be expressed as two two-point correlation functions:

\begin{align}
\xi_{+/-}^{ij}(\theta) &=& (1+m^i)(1+m^j)\int \frac{dl\,l}{2\pi}\,J_{0/4}(l\theta) \vs
&& \hspace{-0.5cm}\int d\chi  \frac{q_\kappa^i(\chi)q_\kappa^j(\chi)}{\chi^2}P_{\mathrm{NL}}\left(\frac{l+1/2}{\chi},z(\chi)\right),
\end{align}
where $m^{i}$ is the multiplicative bias of the shear estimates in the $i$th tomographic bin, $J_{0/4}(l\theta)$ are $0$th and $4$th order Bessel functions. The lensing kernel, $q$, is given by

\begin{align}
q_\kappa^{i}(\chi) = \frac{3 H_0^2 \Omega_m }{2 \mathrm{c}^2}\frac{\chi}{a(\chi)}\int_\chi^{\chi_{\mathrm{h}}} \mathrm{d} \chi' \frac{n_{\kappa}^{i} (z(\chi')) dz/d\chi'}{\bar{n}_{\kappa}^{\
i}} \frac{\chi'-\chi}{\chi'} \,,
\end{align}
with the Hubble constant given as $H_0$, $c$ the speed of light, and $a$
the scale factor. Tangential shear--galaxy cross correlations, 
often referred to as galaxy--galaxy lensing, can be expressed as:

\begin{align}
  \gamma_{\mathrm t}^{ij}(\theta) &=& (1+m^j)\int
  \frac{dl\,l}{2\pi}\,J_2(l\theta)\, \int\!\! d\chi\!
  \frac{q_{\delta_{\mathrm{g}}}^i\!\!\left(\frac{l+1/2}{\chi},\chi\right)
    q_\kappa^j(\chi)}{\chi^2} \vs && \, \times
  P_{\mathrm{NL}}\!\left(\frac{l+1/2}{\chi},z(\chi)\right).
\end{align}
We have included multiplicative bias in these expressions for
completeness only, as we do not model this in these simulations. 
Again, we see that galaxy bias and the non-linear matter power
spectrum are key ingredients in these observables and thus must be 
accurately modeled in our simulations. In principle, baryons affect
the matter power spectrum at small scales, but the DES cosmology
analyses have made conservative scale cuts in order to mitigate 
these effects.

\subsection{Cluster Counts}
Finally, we wish to model the number densities of clusters as 
a function of richness, $\lambda$, an observable that is
tightly correlated with the dark matter halo mass of these
clusters, and redshift:

\begin{equation}
\label{eqn:cluster_counts}
 n(\Delta \lambda_i,z)= \int_{0}^{\infty} \de M\ n(M,z) 
 \int_{\Delta \lambda{\rm i}} \de \lambda p(\lambda | M,z) \, ,
\end{equation}

where $\Delta \lambda_i$ represents a bin in richness, $n(M,z)$ is the number density 
of halos as a function of mass and redshift, and $p(\lambda | M,z)$ is the 
probability that a halo of mass $M$ has observed richness $\lambda$, also known
as the mass--richness relation, or mass-observable relation (MOR). 
Accurate modeling of the halo mass function and mass--richness relation 
are necessary in order to reproduce observed cluster number counts.

\section{Creating Synthetic Dark Energy Surveys}
\label{sec:algorithm}

A brief summary of the algorithm steps are as follows:
\begin{enumerate}
\item Determine matter distribution (\S \ref{sec:nbody}):
\begin{enumerate}
\item run N-body simulations, output lightcones for large-volume simulations
and snapshots for high-resolution simulations
\item find dark matter halos
\item run merger tree on high-resolution simulations
\item calculate densities on halo centers; calculate densities
on particles for large volume boxes
\end{enumerate}
\item Add galaxies (\S \ref{sec:galaxies}):
\begin{enumerate}
\item calibrate luminosity-density relation in the SDSS $z=0.1$ frame $r$-band
on a high-resolution tuning simulation using abundance matching to predict the galaxy distribution
\item add galaxies to large volume lightcones, based on luminosity-density relation
\item measure the observed distribution of SEDs at a given luminosity and
  galaxy density in SDSS and use to assign SEDs to simulated galaxies
\end{enumerate}
\item Lens galaxies (\S \ref{sec:pofz_cosmicshear}):
\begin{enumerate}
\item add unlensed galaxy shapes and sizes
\item calculate lensing fields (shear, deflection, convergence, rotation) via ray-tracing at all galaxy positions
\item lens (magnify and distort) galaxies, including multiple images
\end{enumerate}
\item ``Observe'' galaxies (\S \ref{sec:obs_eff} and \S \ref{sec:pofz_cosmicshear}):
\begin{enumerate}
\item rotate into DES footprint, apply survey mask
\item apply photometric errors
\item calculate photometric redshifts
\item select samples
\end{enumerate}
\end{enumerate}

The final result is a synthetic galaxy catalog containing positional, spectroscopic, and photometric information, as well as additional properties such as
photometric redshift information.  In addition, the catalog contains
measured properties of the matter distribution, including halo
properties and weak-lensing quantities.  Fig. \ref{fig:buzzard_y1_map} shows an
example of one of our simulations, displaying the underlying projected matter distribution, an observed cluster catalog and weak-lensing shear around a massive cluster. A full list of modeled galaxy properties appears in Appendix
\ref{app:tags}.

\begin{deluxetable*}{cccccccccc}
  \tabletypesize{\footnotesize}
  \tablecaption{
  Description of the simulations used to create the particle lightcone. \label{table:simulations}}
  \startdata
  \\
  \hline
  Name & $z_{min}$ & $z_{max}$ & $L_{box}$ & $N_{part}$ & $m_{part}$
  & $\epsilon_{Plummer}$
  & $M_{r, {\rm lim}}$ &
  N$_{\rm halos}$ w/
  50 particles & N$_{\rm galaxies}$ to DES limit\\
  \hline
  \hline
  T & tuning only & tuning only & 400 \hmpc & 2048$^3$& 4.8 $\times 10^{8} \hinv\msun $& 5.5 \hkpc& -- & 5.3M & NA\\
  L1 & 0.0 & 0.32 & 1.05 \hGpc & 1400$^3$ & 3.3$\times 10^{10} \hinv\msun $& 20 \hkpc & -10 & 4.7M  & $1 \times 10^8$\\
  L2 & 0.32 & 0.84 & 2.6 \hGpc &  2048$^3$& 1.6$\times 10^{11} \hinv\msun $&35 \hkpc & -16.6 & 8.2M & $3 \times 10^8$\\
  L3 & 0.84 & 2.35 & 4.0 \hGpc & 2048$^3$& 5.9$\times 10^{11} \hinv\msun $& 53 \hkpc & -19.1& 1.4M & $3 \times 10^8$\\
  \enddata
\end{deluxetable*}

\subsection{N-body Simulation Methodology}
\label{sec:nbody}
The key details of the simulations run are listed in
Table~\ref{table:simulations}. The simulations L1, L2, and L3 are combined
to build the particle lightcone that generates 10,313 square degrees of unique,
contiguous sky. The box T is used to tune the galaxy assignment algorithm as
described below and thus only one per cosmological model is needed. Note that
at higher redshifts in a flux-limited survey, the smallest halo mass needed 
to model a given set of galaxies increases, since progressively higher 
luminosity galaxies living in more massive halos are probed at higher
redshifts. Thus, using simulation volumes of progressively lower resolution
as a function of redshift in the lightcone allows us to lower the
computational cost of the simulations. The disadvantage of this technique
is that it leaves discontinuities in cosmic structures along the line-of-sight
at the edges between the different lightcones. We have placed the transitions in
redshift where typical red sequence galaxy photometric redshifts have worse 
performance due the \Abreak moving between filters \citep{rykoff_etal:13}.

Briefly, we use \textsc{2LPTic} \citep{crocce_etal:06} and \textsc{L-Gadget2} \citep{springel_etal:05} to initialize and run our simulations. We use a
\LCDM\ cosmology with $\Omega_{m}=0.286$, $h=0.7$, $\sigma_{8}=0.82$, $n_s=0.96$,
and $\Omega_b=0.046$, with three massless neutrino species and $N_{eff}=3.046$. 
We refer to this as the Buzzard cosmology, hence the name of the simulation suite.
We have made further specialized modifications to the codes to initialize 
simulations of generic dark energy models using second-order Lagrangian
perturbation theory and to generate lightcone outputs on-the-fly. We find
halos with \textsc{rockstar} \citep{behroozi_etal:13a}
and generate merger trees on our high resolution simulation with 
\textsc{consistent-trees} \citep{behroozi_etal:13b}. Finally, we use
\textsc{calclens} \citep{becker:12} to compute the weak-lensing shear,
magnification and lensed galaxy positions from the lightcone outputs.
The full details of our N-body and simulation post-processing methodology are
described in Appendices~\ref{app:nbody}, \ref{app:weaklens},
and \ref{app:de2lptic}.

\subsection{N-body simulation validation}

\begin{figure*}
  \centering
  \includegraphics[width=0.98\columnwidth]{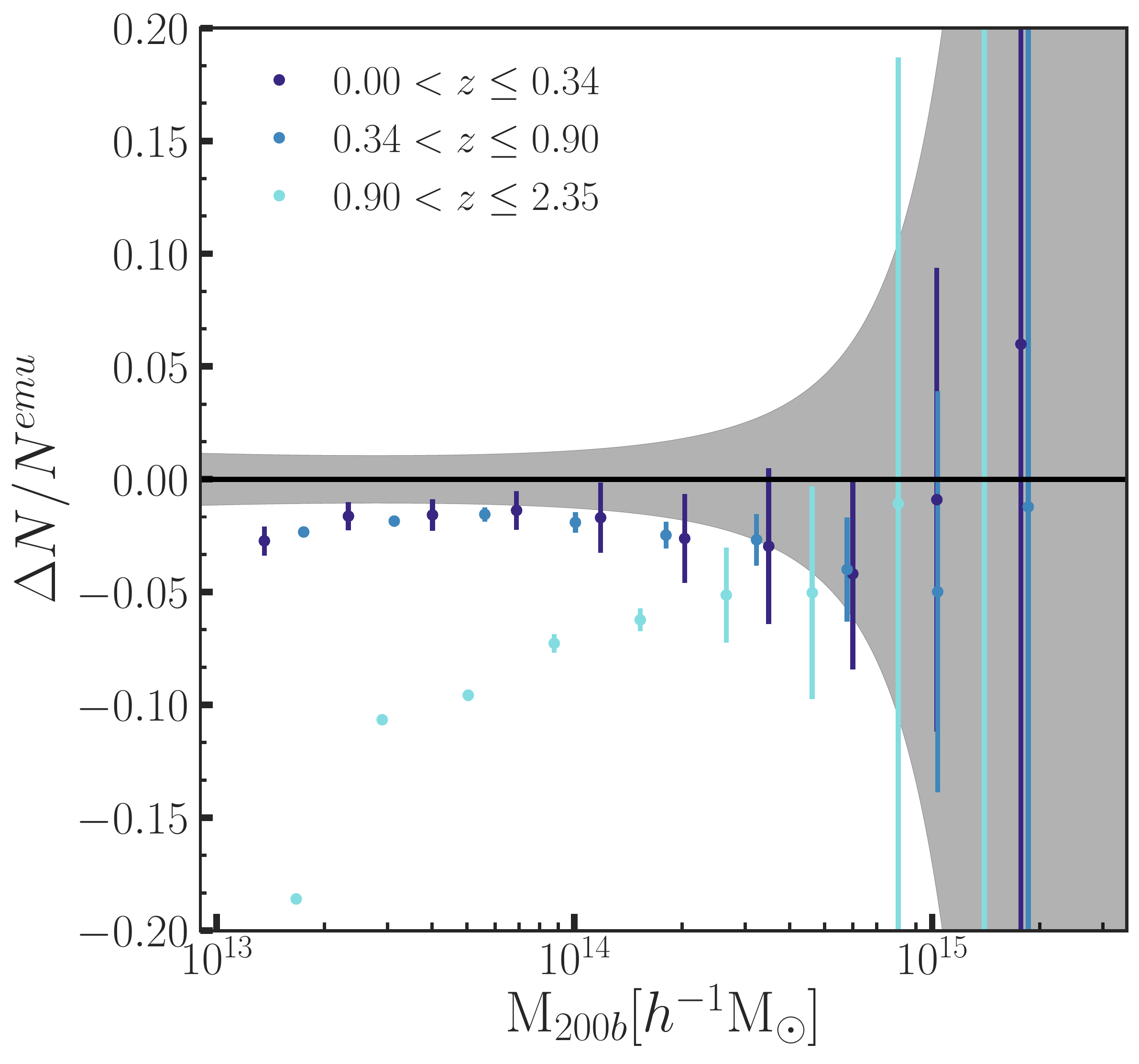}
  \includegraphics[width=\columnwidth]{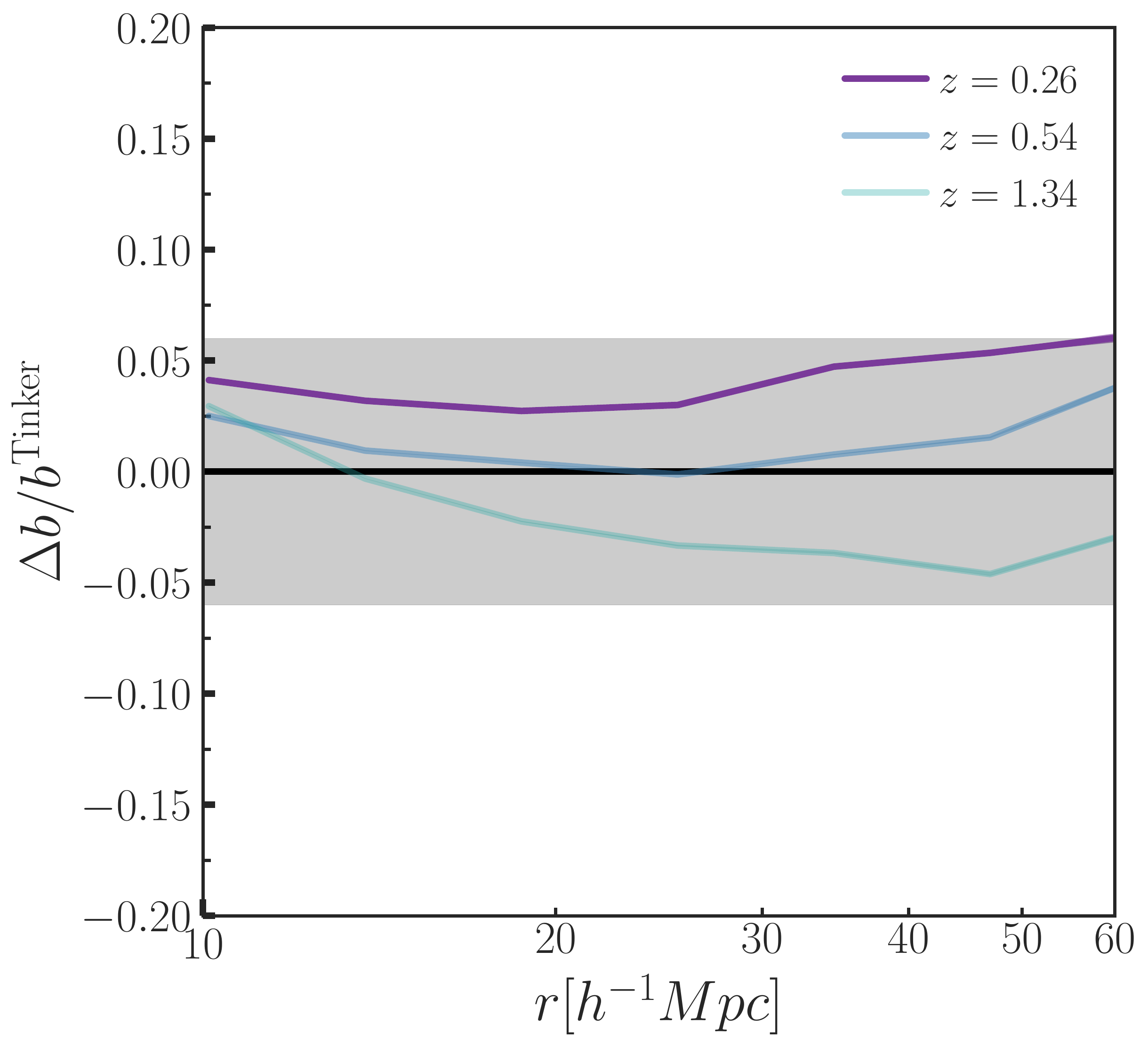}
  \caption{\textit{Left}: The fractional error in the halo mass
    function measured from the mean of our simulations with respect to
    the \citet{McClintock2018} halo mass function emulator. The three redshift
    bins correspond to the \textsc{L1}, \textsc{L2} and \textsc{L3} simulations
    from low to high redshift. The grey band represents the accuracy requirement
    for DES Y5 at z=0.5 as computed in \citet{McClintock2018}. The accuracy of
    the emulator is better than this at all masses. The discrepancies at $z<0.9$
    are likely due to differences in halo definition 
    (see discussion in Appendix \ref{sec:halofinding}). 
    The discrepancies seen at high redshift, where the emulator over-predicts the 
    simulations, are due to resolution effects in the \textsc{L3} lightcone.
    \textit{Right}: Fractional error of halo bias measured in a bin 
    with mean mass of $4\times 10^{13} \hmsun$ for three different redshifts
    (lines) with respect to the predictions from \citet{Tinker2010}.
    The measurements are averaged over all $L1$, $L2$ and $L3$ simulations for the first, second and
    third redshift bin respectively. The grey band represents the quoted 6\% error
    in the \citet{Tinker2010} predictions.}
  \label{fig:halo_comp}
\end{figure*}

\begin{figure*}
  \centering
  \includegraphics[width=\columnwidth]{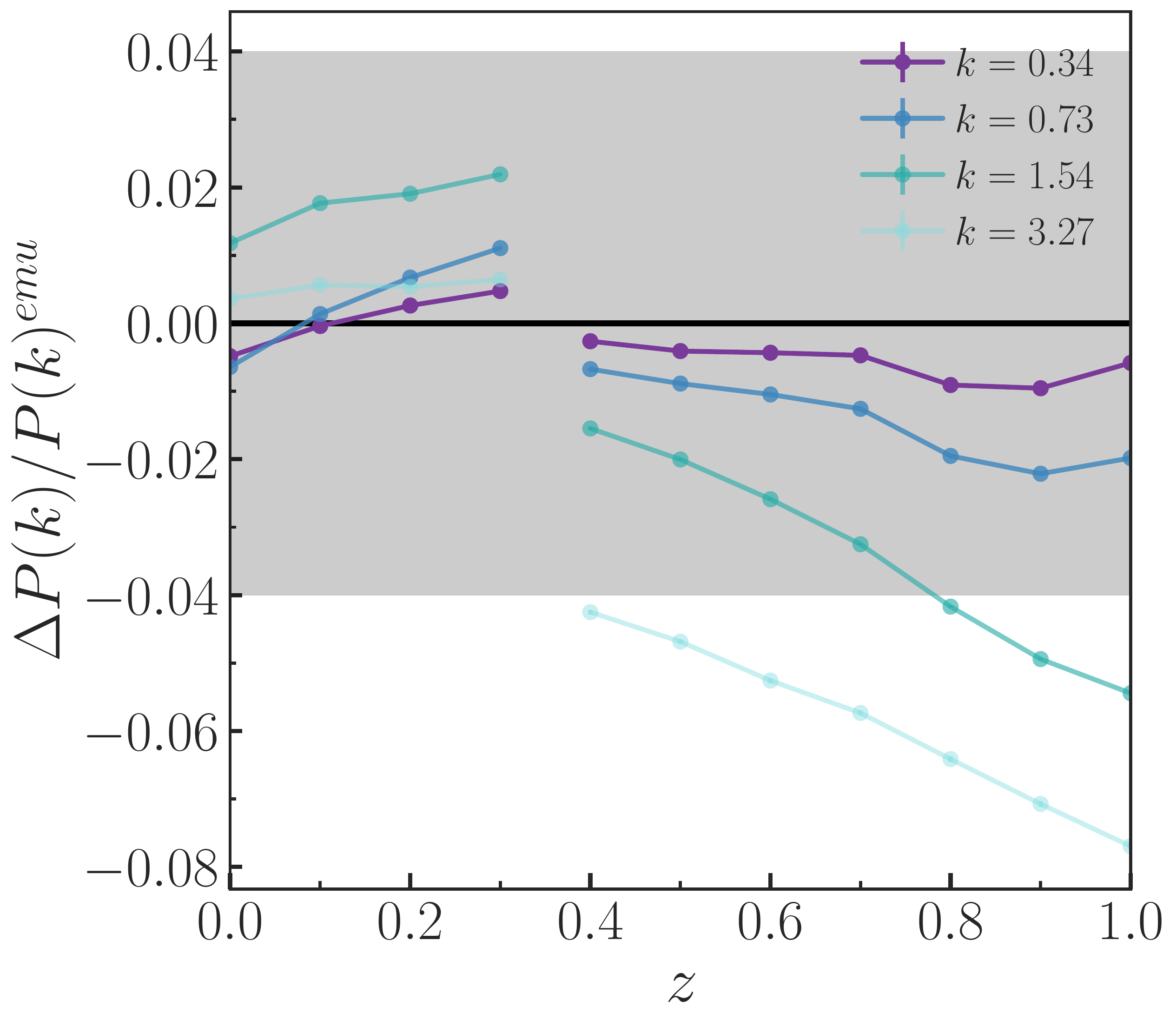}
  \includegraphics[width=0.92\columnwidth]{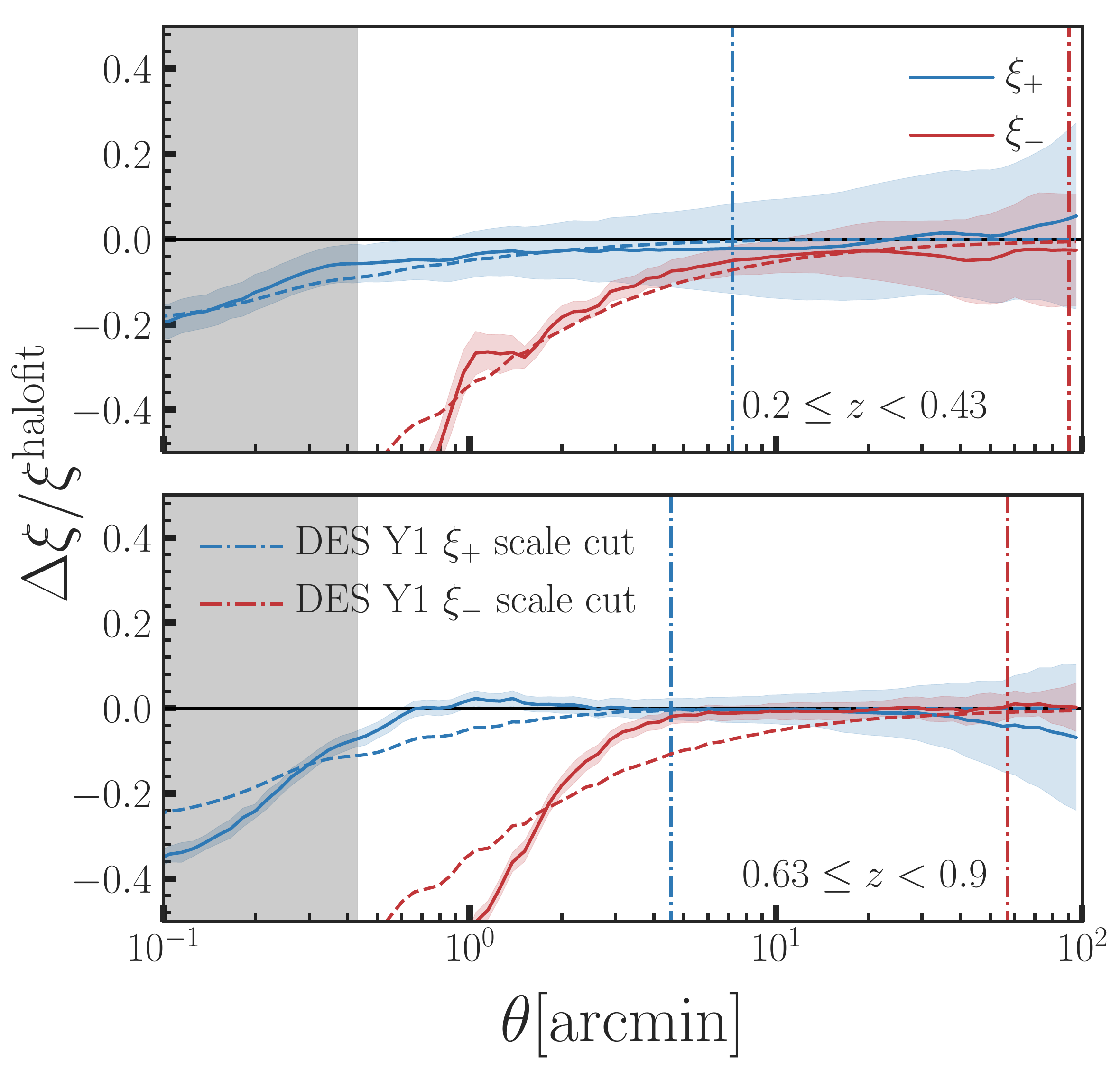}
  \caption{ 
    \textit{Left}: Ratio of the the matter power
    spectrum measured in snapshots of the $L1$ (for $z<0.35$) and $L2$ (for
    $z>0.35$) simulations to the CosmicEmu \citep{Lawrence2017} as a function of redshift 
    for different values of $k$. $L3$, not shown, is used for $z>0.9$ and is lower 
    resolution than $L2$. The switch between $L1$ and $L2$ is marked by a break in 
    the lines.
    \textit{Right}: Comparison of $\xi_{+/-}$ (blue/red)
    without shape noise (solid) to halofit (dashed) for two different redshift bins
    averaged over 18 Y1 footprints (top panel). The fractional deviation from 
    halofit \citep{takahashi2012} (dashed) compared to the prediction made 
    using a halofit power spectrum truncated at a fixed $\ell$ approximately
    corresponding to the ray-tracing resolution is shown in the bottom panel. 
    The different scale dependence of the resolution effects in the two redshift 
    bins indicates that they are due to the resolution of the underlying N-body 
    simulation and not the ray-tracing. The gray region corresponds to the 
    approximate angular resolution of the HEALPix grid used to perform the
    ray-tracing.}
  \label{fig:nbody_comp}
\end{figure*}

We have compared our simulations to standard fitting functions and analytic
approximations, the results of which are shown in Figs.~\ref{fig:halo_comp} and
~\ref{fig:nbody_comp}. As seen in Fig.~\ref{fig:nbody_comp}, the power spectra, $P(k)$, in both the L1 and L2 boxes agree with the predictions from 
the \textsc{CosmicEmu} emulator within their quoted errors \citep[4\%;][]{Lawrence2017} 
for wavenumbers $k<2~\hmpc$(left panel). We see poor agreement above scales of
$k=2~\hmpc$ in the L2 box. The matter in the L3 box is weighted significantly 
less by the lensing kernel of typical DES source galaxies, making its
contribution to weak-lensing observables small, and so is not shown. However,
as discussed in Appendix~\ref{app:weaklens} and demonstrated in the right
panel of Fig.~\ref{fig:nbody_comp}, the weak-lensing signals from our 
simulations, presented here in the form of $\xi_{+/-}$ measured without shape
noise and averaged over all 18 simulations, are affected by the relatively 
lower resolution of the L2 and L3 boxes. 

Similarly, we find that the L1 and L2 boxes agree with halo mass function
and halo bias predictions in the literature \citep{McClintock2018, Tinker2010}, 
as seen in the left and right panels of Fig.~\ref{fig:halo_comp} respectively. 
The $\sim 1\%$ deviations seen in the mass function are due the differences
in halo definition used in our simulations and that used in \citet{McClintock2018}
as described in Appendix \ref{sec:halofinding}. The halo mass function of L3 is
again affected by the lower resolution of this box, but the vast majority of 
clusters detected in DES have redshifts $z<0.9$, as the red sequence used to 
find clusters redshifts out of the DES bands at this point, and more generally
becomes less well defined above $z=1$. Thus, the impact of resolution effects
in \textsc{L3} on photometrically detected cluster observables is negligible, although
it may be important for cluster selections that are less redshift dependent such
as those based on the thermal Sunyaev-Zel'dovich effect \citep[e.g.][]{Bleem2014}. 

Both the resolution effects and the discontinuities in the matter distribution
were shown to have negligible impact on inferring the true cosmology of these
simulations using a $3\times2$-point analysis with a DES Y1 covariance matrix in 
\citet{MacCrann17}. This statement is analysis dependent, and analyses that use
smaller scales, or that have smaller errors on their observables, such as DES Y3
and Y5 analyses, may not be immune to systematic effects due to the compromises
made in creating these simulations. See Appendices~\ref{app:nbody} and
\ref{app:weaklens} for additional discussion.

\subsection{Building Galaxy Populations: ADDGALS and the Color-Density Relationship}
\label{sec:galaxies}
We use the \addgals\ algorithm to populate our lightcones
with galaxies. We briefly describe the algorithm here and refer the reader to 
\citet{wechsler_etal:19} for more details. Additional implementation details related 
to high redshift extensions of the \textsc{Addgals} model are presented in Appendix \ref{app:gals}.

\textsc{Addgals} uses three main distributions to populate galaxies: 
\begin{itemize}
\item A luminosity function, $\phi(M_{r},z)$
\item The distribution of central galaxy absolute magnitude at fixed halo mass and redshift, $P(M_{\rm r,cen}|M_{\rm vir},z)$
\item The distribution of galaxy overdensities conditioned on absolute magnitude and redshift, $P(R_{\delta}|M_{r}, z)$, where $R_\delta$ is the radius around each galaxy enclosing $1.3\times10^{13}\,\hmsun$
\end{itemize}
These PDFs are measured by applying a SHAM model to the T1 simulation, using a luminosity function, $\phi(M_{r},z)$, determined from data as described below. In order to determine $P(M_{\rm r,cen}|M_{\rm vir},z)$ we fit a functional form to the mean relation in the SHAM and assume Gaussian scatter around that, i.e. 
\begin{equation}
P(M_{\rm r,cen} | M_{\rm vir}) = \mathcal{N}(M_{\rm r,cen}(M_{\rm vir}), \sigma_{M_{\rm r,cen}}).
\end{equation}
$P(R_{\delta}|M_{r}, z)$ is determined by measuring $R_{\delta}$ around each simulated galaxy in the SHAM catalog and measuring the distribution of $R_\delta$ for galaxies above a given magnitude cut, $M_r$, in a snapshot with redshift $z$. A functional form is then fit to this distribution as a function of $M_r$ and $z$.

We use the best-fit SHAM model from \citet{Lehmann2017}, which postulates, roughly, that brighter galaxies live in halos with higher peak circular velocities, while allowing for some scatter in the $v_{peak}$ - luminosity relation. It has been shown to reproduce the luminosity dependent clustering as measured in SDSS to high precision \citep[e.g.][]{conroy_etal:06,Reddick12,Lehmann2017}. For more details about the SHAM model 
that we use, and how $P(R_{\delta}|M_{r}, z)$ is determined we direct the reader to \citet{wechsler_etal:19}.

In order to assign galaxies to our lightcones, we first assign central galaxies to resolved halos using the $P(M_{\rm r, cen}|M_{vir},z)$ defined by the SHAM. Doing so only accounts for a small fraction of the galaxies observed by a survey such as DES. To populate our lightcones with the remaining galaxies, we use the $P(R_{\delta}|M_{r}, z)$, drawing a galaxy from the observed luminosity function and an overdensity for this galaxy from $P(R_{\delta}|M_{r}, z)$, and assigning it to a dark matter particle with the correct overdensity $R_\delta$. This procedure results in a galaxy catalog that matches the scale- and luminosity-dependent two-point clustering in a single magnitude band. We choose to assign galaxies to particles as they are convenient points around which to measure densities, but in principle we can place galaxies anywhere the local densities required for the method can be measured. Using particles places a limit on the number density of galaxies that we can assign, but the catalogs here do not approach this limit.

The above procedure results in a catalog with a single absolute magnitude per galaxy. We also wish to assign SEDs to each galaxy. Doing so involves two additional observational inputs:
\begin{itemize}
\item A training set of galaxies from which to draw SEDs
\item The fraction of galaxies that are red at fixed $M_r$ and $z$
\end{itemize}
We assign SEDs from the SDSS DR7 VAGC \citep{blanton05} to our simulated galaxies by imposing that our simulation match the SED--luminosity--density relationship measured in the SDSS data. The evolution in this relationship is calibrated to the evolution of the red fraction of galaxies from PRIMUS \citep{primus} as a function of $M_r$ and $z$. The relatively small area of PRIMUS, 9 sq. degrees, may contribute non-negligible sample variance to this relation, but its unparalleled spectroscopic depth of $r\sim 23.5$ makes it ideal for the purposes of this calibration.

In more detail, the assignment of galaxy SEDs proceeds by measuring the projected distance to the fifth nearest neighbor, $\Sigma_5$, in a small bin in redshift around each galaxy in both the the SDSS DR7 VAGC and our mock catalog. We sort these distances producing a \textit{rank} $R_{\Sigma_5}$. We then assign SEDs to each mock galaxy by selecting a galaxy from the data in the same bin of absolute magnitude $M_r$ and $R_{\Sigma_5}$. Evolution in the SED--luminosity--density relation is accounted for by preferentially drawing from blue galaxies over red ones at higher redshifts, enforcing the constraint that the red fraction of galaxies match that measured in PRIMUS as a function of luminosity and redshift. The assumptions made by this method do not hold in detail, and the resulting imperfections are discussed below. SEDs are represented as weighted sums of 5 \textsc{Kcorrect} \citep{blanton_etal:03kcorr} templates, allowing for efficient computation of the intrinsic observed magnitudes of each mock galaxy in a variety of pass bands. See Appendix~\ref{sec:colors} for more details.

The above procedure relies on a measurement of the galaxy luminosity function in the data over the
range of redshifts of interest. This task is non-trivial given the large number of systematic 
effects in the measurement of galaxy magnitudes, survey incompleteness, redshift errors, and other systematics.
Further, we would like our luminosity function, when integrated over a typical $\Lambda$CDM volume, 
to match the observed DES Y1 galaxy counts as a function of magnitude. To complete this task, we
start with a luminosity function as measured with small statistical error at low redshift using the 
method described in \cite{Reddick12} based on the SDSS spectroscopic sample. To account for 
redshift evolution we fit for an additive evolution in $M_{*}$ by first populating a full lightcone
using our fiducial luminosity function and then minimizing the difference between our fiducial 
counts in the DES Y1 bands and the observed counts in the $\approx 1.5 ~\textrm{deg}^2$ overlap 
between DES and COSMOS, which is approximately 1 magnitude deeper than the wide field observations.
This correction is described in more detail in Appendix~\ref{sec:lf}.

\begin{figure}
  \includegraphics[width=\columnwidth]{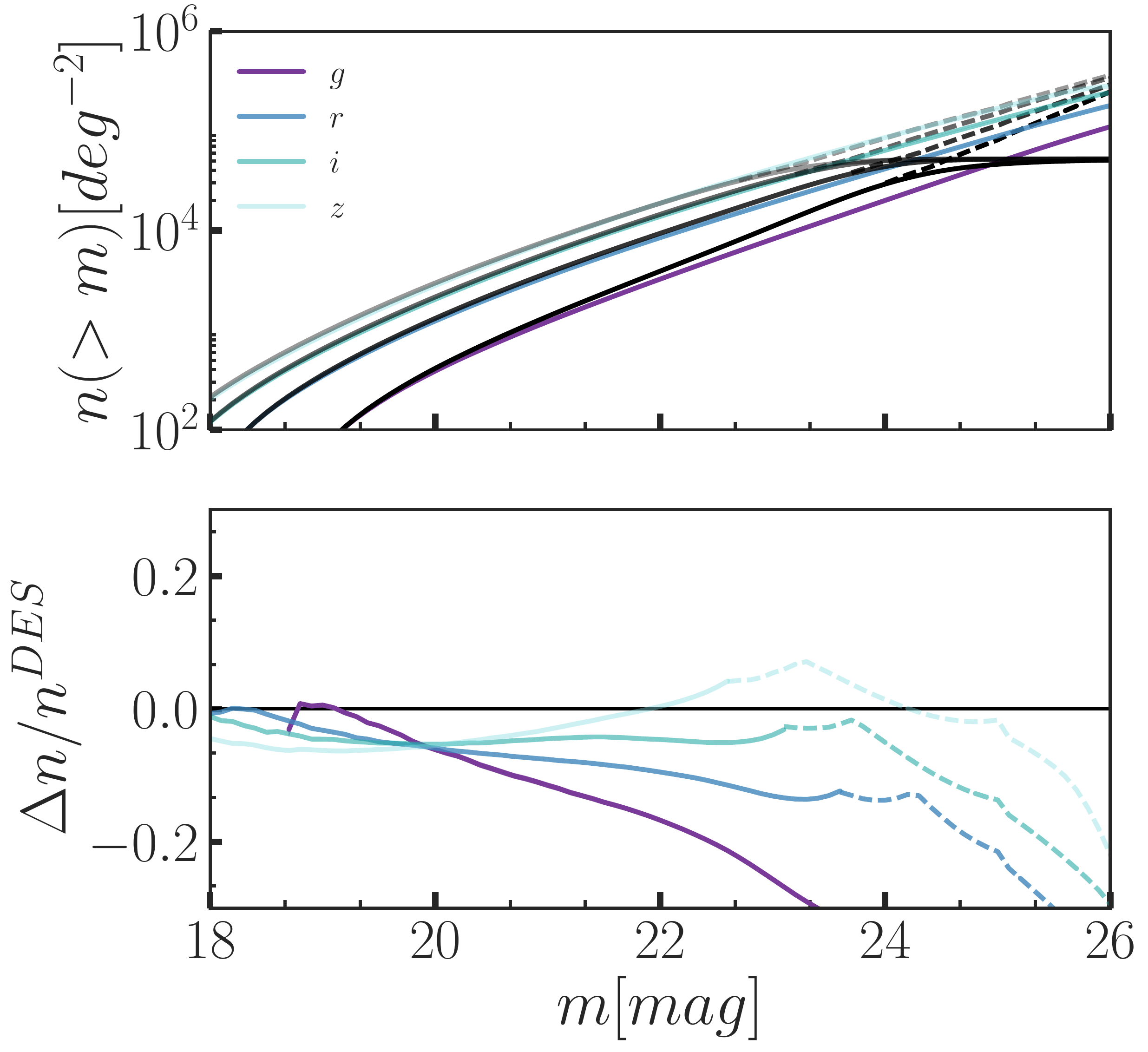}
  \caption{\textit{(top)} Cumulative number of observed galaxies as
  a function of true magnitude per square degree in buzzard (colors)
  compared to the same quantity as a function of observed magnitude
  in the DES Y1 wide field data (solid black) and a power law 
  extrapolation to fainter magnitudes (dashed black). 
  \textit{(bottom)} Fractional deviation of the simulations from
  the data and its power law extrapolation with Poisson error bars
  which are mostly not visible due to their small amplitude. At 
  magnitudes fainter than $r=24$, we expect $>10\%$ deficits
  in Buzzard due to the absence of galaxies with $z>2.35$.}
  \label{fig:gals}
\end{figure}

\begin{figure*}
  \includegraphics[width=\linewidth]{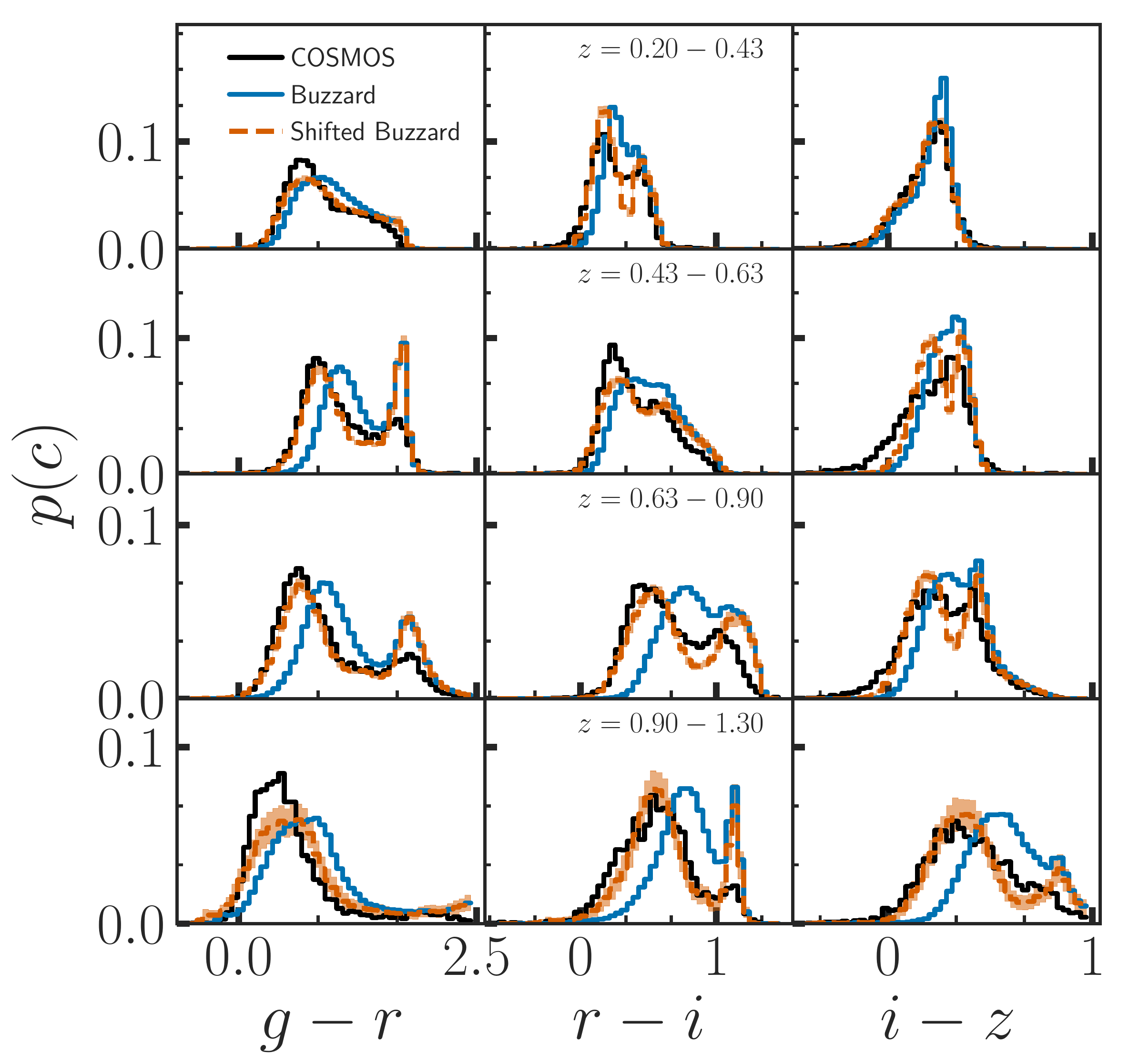}
  \caption{Comparison of the $g-r$, $r-i$ and $i-z$ colors between
  Buzzard and COSMOS for magnitudes $18<m_{r}<23$ in the redshift
  bins used for source galaxies in the Y1 $3\times2$ point analysis.
  Dashed orange lines represent Buzzard colors where we have shifted
  the mean of the blue sequence in each redshift bin to match the 
  observed colors. The significant improvement in the agreement
  between the shifted Buzzard distribution and COSMOS shows that the
  most relevant difference between COSMOS colors and Buzzard is just
  such a shift. Error bars are estimated via jackknife from 128
  COSMOS sized patches in Buzzard.}
  \label{fig:colors}
\end{figure*}

\begin{figure*}
\centering
  \includegraphics[width=\columnwidth]{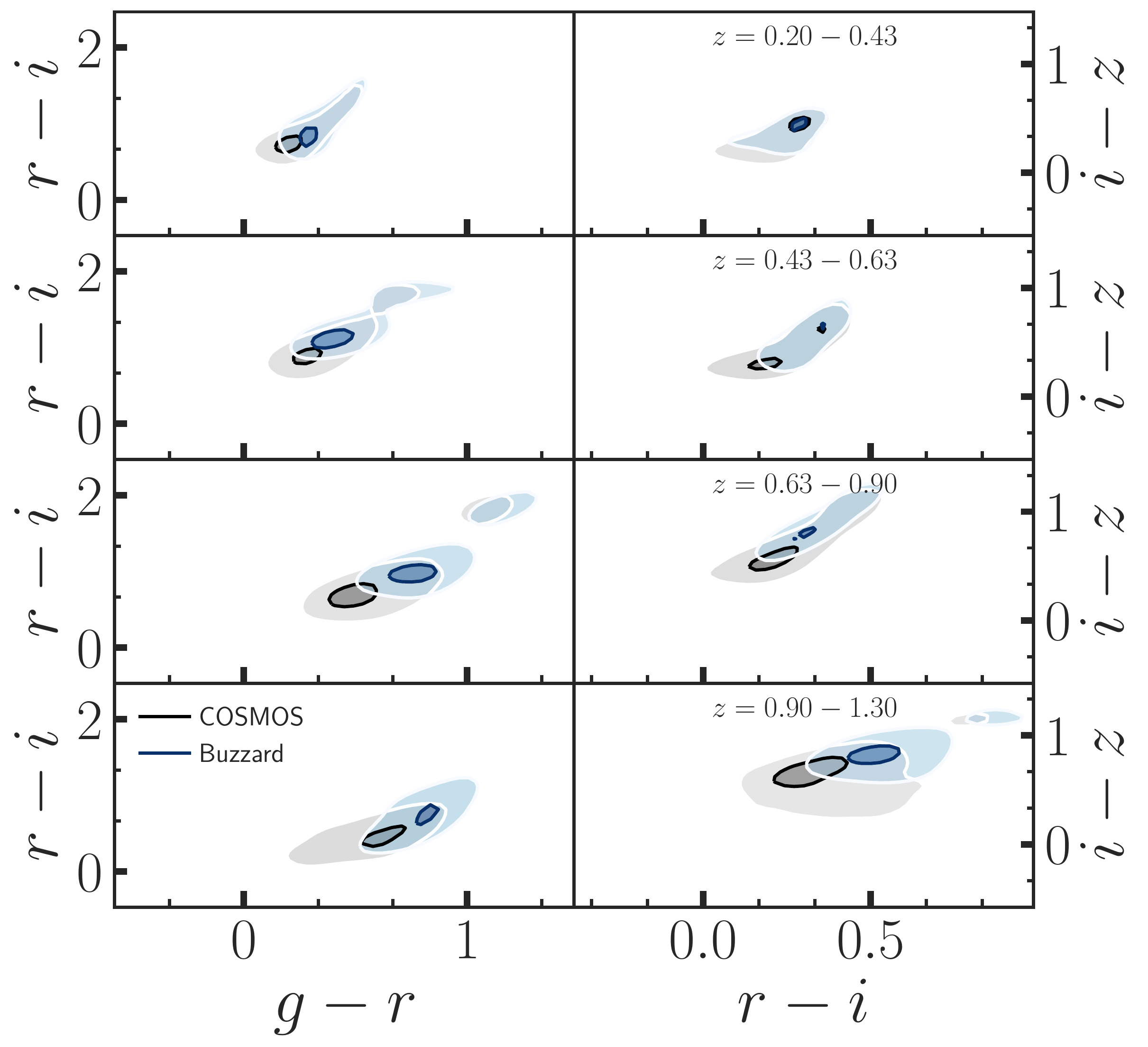}
  \includegraphics[width=\columnwidth]{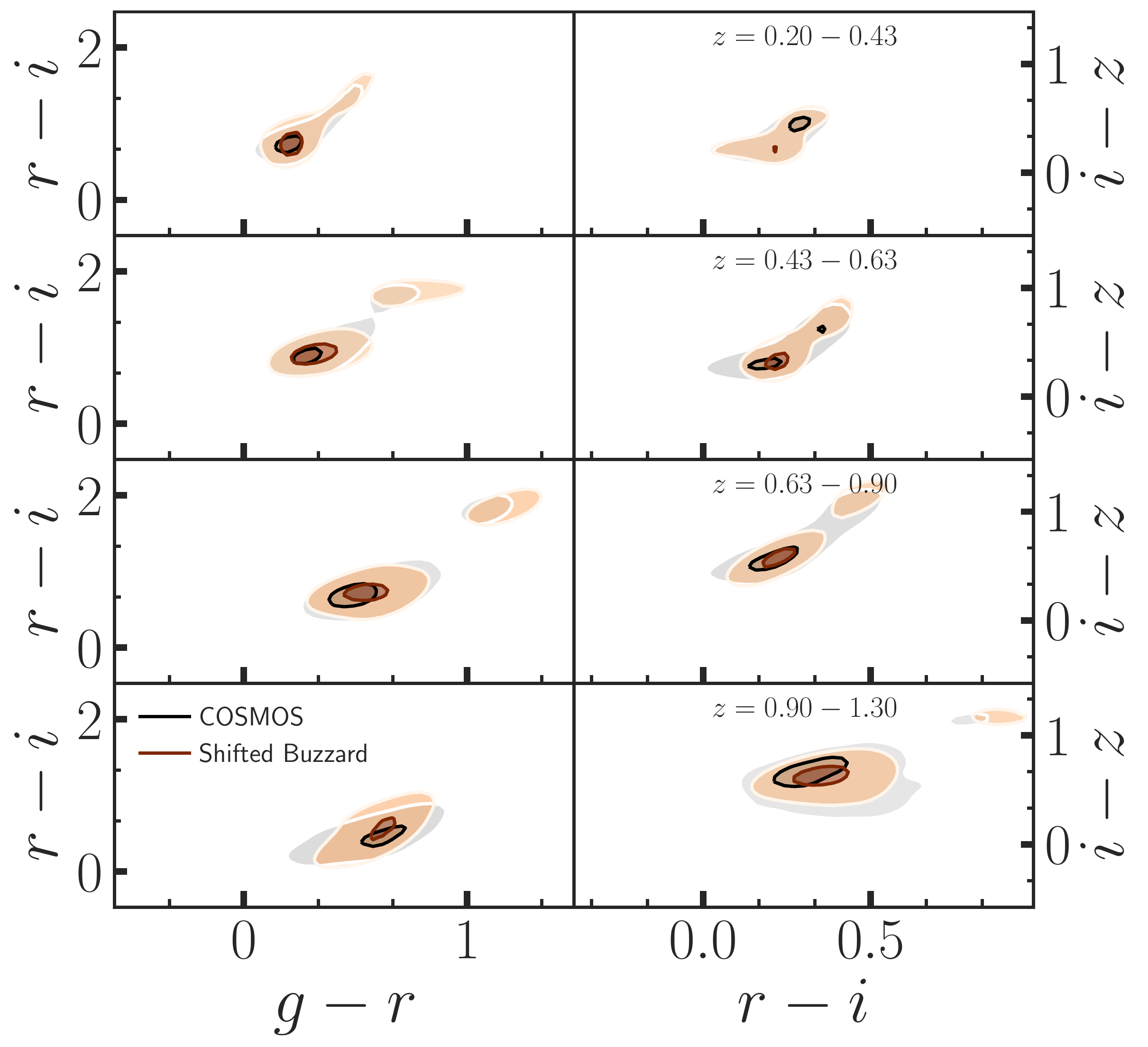}  
  \caption{\textit{(left)} Comparison of two dimensional color distributions between Buzzard and COSMOS for magnitudes $18<m_{r}<23$ in the redshift bins used for source galaxies in the Y1 $3\times2$ point analysis. The left column is the joint distribution of $g-r$ and $r-i$ colors and the right is $r-i$ and $i-z$. Contours represent the one and two sigma boundaries of the distributions. \textit{(right)} The same, but for Buzzard colors with the blue cloud shifted in the same way as in Fig. \ref{fig:colors}. Again, the significant improvement in the agreement between the shifted Buzzard distribution and COSMOS shows that the most relevant difference between COSMOS colors and Buzzard is just such a shift.}
  \label{fig:color-color}
\end{figure*}

The final magnitude counts and the color distributions of our mock galaxies compared to the DES Y1 
data are shown in Figs.~\ref{fig:gals}, ~\ref{fig:colors} and ~\ref{fig:color-color}. We find that
our catalogs are in agreement with measurements of DES counts to $\sim 10 - 20 \%$ accuracy 
depending on the band as can be seen in Fig. \ref{fig:gals}, and are roughly consistent
with a power-law extrapolation to fainter magnitudes, represented by the dashed lines in
that figure. At $i=24$ we expect $\sim 10\%$ under-predictions in counts
due to the fact that we are not populating galaxies with $z>2.35$, with the deficit becoming 
worse at fainter magnitudes. Residual discrepancies between the different bands
are due to the fact that colors in these simulations become redder than what is
observed in the data at high redshift (see the discussion below), leading to relative
underestimates of $\vec{n}(>\vec{m})$ for bluer bands. 
The regime brighter than shown in Fig.~\ref{fig:gals} matches counts in the SDSS 
main sample to $\sim 10\%$ as shown in \citet{wechsler_etal:19}. These 
deviations are small enough to allow for the generation of photometric 
cluster and LRG samples that provide reasonable facsimiles of a number of statistics
as measured in the DES Y1 data as discussed in \ref{sec:redmagic} and \ref{sec:redmapper}.

The Figs.~\ref{fig:colors} and \ref{fig:color-color} show the distribution of colors in Buzzard compared to those measured in the DES overlap with COSMOS in a magnitude bin of $18<m_{r}<23$. Brighter than this, our colors are well validated by SDSS, and fainter than this, the COSMOS overlap with DES has large photometric noise and is likely incomplete. We bin the COSMOS galaxies
by redshift using their \textsc{BPZ} photometric redshifts which have significantly smaller dispersion than the size of the redshift bins used here \citep{Laigle2016}. At low redshift, the colors show relatively small deviations from those in the data, a non-trivial accomplishment 
given that the apparent magnitude range probed here is significantly deeper than that used in our training set ($m_{r}<17.7$). 
At higher redshifts, we see two main modes of deviation between COSMOS and Buzzard. 
The first is that the mean of the blue sequence of galaxies is significantly bluer
in the data than in the simulations. To demonstrate this we have also plotted Buzzard
colors where we have shifted the mean of the blue sequence in each redshift bin separately for each color to match the COSMOS data. Error bars on the shifted distribution in Fig.~\ref{fig:colors} are calculated via jackknife on 128 COSMOS sized patches in Buzzard. The agreement between this shifted Buzzard distribution and distribution of colors in COSMOS is much better. This is significant, as it means that improving the colors dramatically in these simulations may be a matter of including relatively minor adjustments to our templates as a function of redshift, rather than incorporating additional templates. Kolmogorov-Smirnov (KS) tests comparing the 1D COSMOS distributions and shifted Buzzard colors still reject the null hypothesis (that the distributions are drawn from the same underlying distribution) at high significance, but the KS statistics improve drastically between the unshifted and shifted distributions in nearly all cases. The other main difference is that at high redshift the width of the red sequence in $g-r$ and $r-i$ is greater in the data than in the simulations. Given that we do not make any effort to evolve our templates using stellar evolution models, it is possible that taking this into effect may resolve some of the discrepancies seen here.

The level of agreement between the simulations and the data in this space has allowed us to run a number of algorithms that rely on reproducing the color--redshift relation in the data, such as photometric redshift and sample selection algorithms. The performance of these on the simulations is discussed below, but in general they produce results which have many of the qualitative features observed in the data, while failing to match in a rigorous statistical sense. Nonetheless, the photometric redshifts and color selected samples in these simulations have proven useful for a number of studies \citep[e.g.][]{MacCrann17}.

To summarize, the free parameters of the model are the following:

\begin{enumerate}
\item The luminosity proxy and scatter assumed in the abundance matching procedure used to tune the \addgals\ method
\item An $r$-band luminosity function and its evolution with redshift
\item The catalog of galaxies from which SEDs are drawn
\item The red fraction of galaxies as a function of absolute magnitude and redshift
\end{enumerate}
and the following data sets are used to tune each of these parameters respectively:

\begin{enumerate}
\item Luminosity-dependent projected clustering in the SDSS DR7 VAGC
\item The luminosity function of the SDSS DR7 VAGC and $\vec{n}(>\vec{m})$ in DES Y1
\item SDSS DR7 color distributions
\item The red-fraction of galaxies as a function of absolute magnitude and redshift in PRIMUS
\end{enumerate}
We have not listed $p(R_{\delta}|M_r,z)$ as a free parameter as it is fully specified by the assumed SHAM model and our fitting procedure.

\subsection{Masking and Observational Effects}
\label{sec:obs_eff}
Once we have populated a lightcone with galaxies and lensed them, we apply
several post-processing steps to approximate the effects of the DES Y1
image processing pipeline, including the effects of masking and the reported
DES Y1 observing conditions. First, we cut out two different kinds of
DES Y1 footprints from catalogs, one (1120 sq. degrees) by excluding $330<\textrm{RA}$ and the S82 region of the footprint, the region of the footprint overlapping with SDSS Stripe 82, and the other by using all area not including the S82 region (1321 sq. degrees). We are able to produce six of the reduced
footprint and two of the full footprint per lightcone simulation. Second, we apply the DES Y1 footprint mask, including all areas with greater than one exposure. We then randomly downsample the galaxies according
the quantity \textsc{FRACGOOD} of the Y1 footprint mask, which describes the amount of masking at scales below the resolution of the masks.
Finally, we use maps of the 10-$\sigma$ limiting magnitude and the effective
exposure time to apply photometric noise to the galaxy magnitudes assuming Poisson sampling statistics to account for errors due to background sky photons as well as photons intrinsic to the galaxy itself. See Appendix~\ref{app:gals} for details.

\section{Comparison to DES Y1 Observations}
\label{sec:valid}

We now compare the suite of mock catalogs to various quantities measured in the
DES Y1 survey. These include photometric redshifts, cosmic shear, galaxy--galaxy
lensing, galaxy clustering, and optically-identified galaxy clusters.

\subsection{Photometric Redshifts and Cosmic Shear}
\label{sec:pofz_cosmicshear}

\begin{figure*}
  \centering
  \includegraphics[width=\columnwidth]{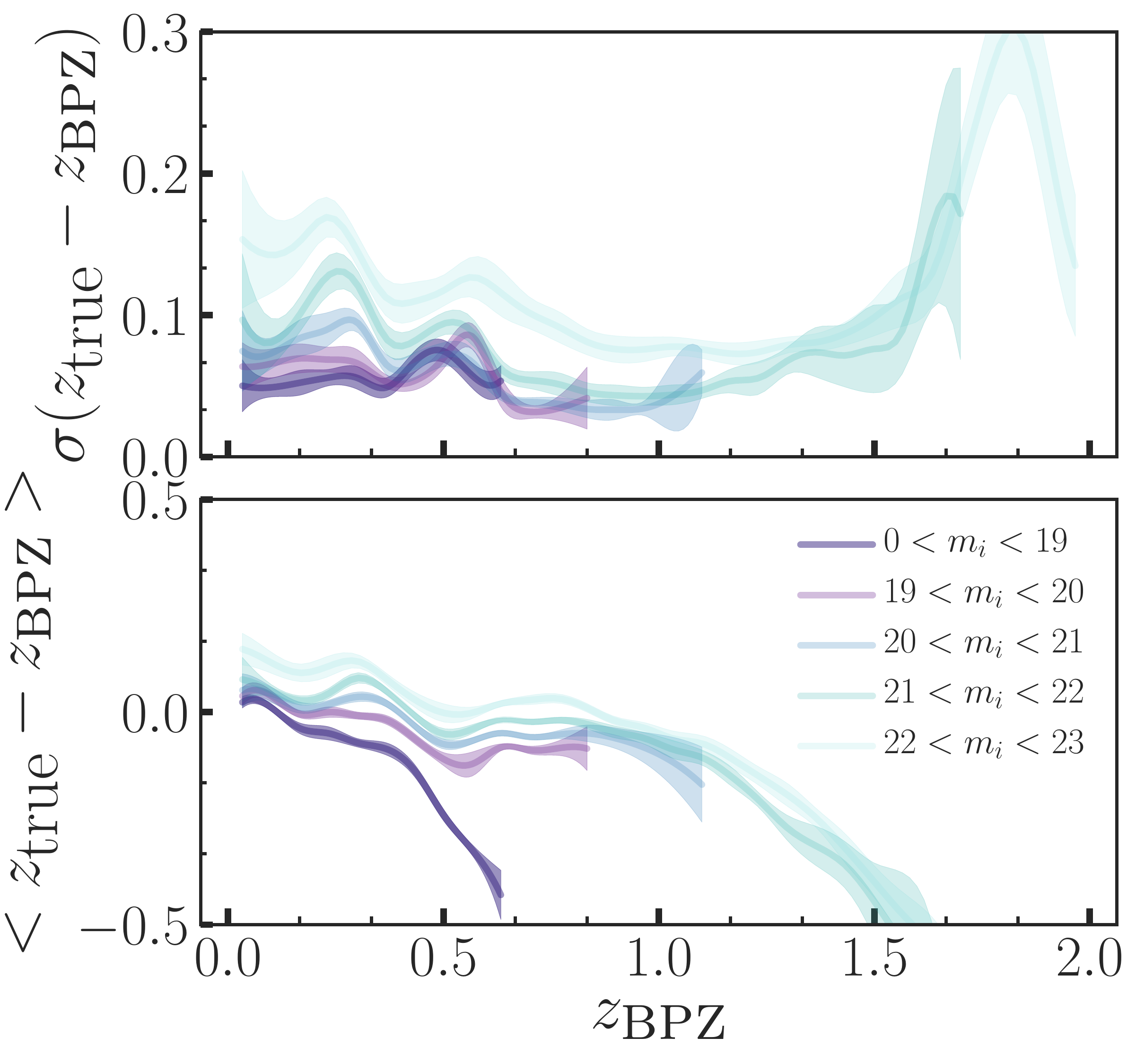}
  \includegraphics[width=\columnwidth]{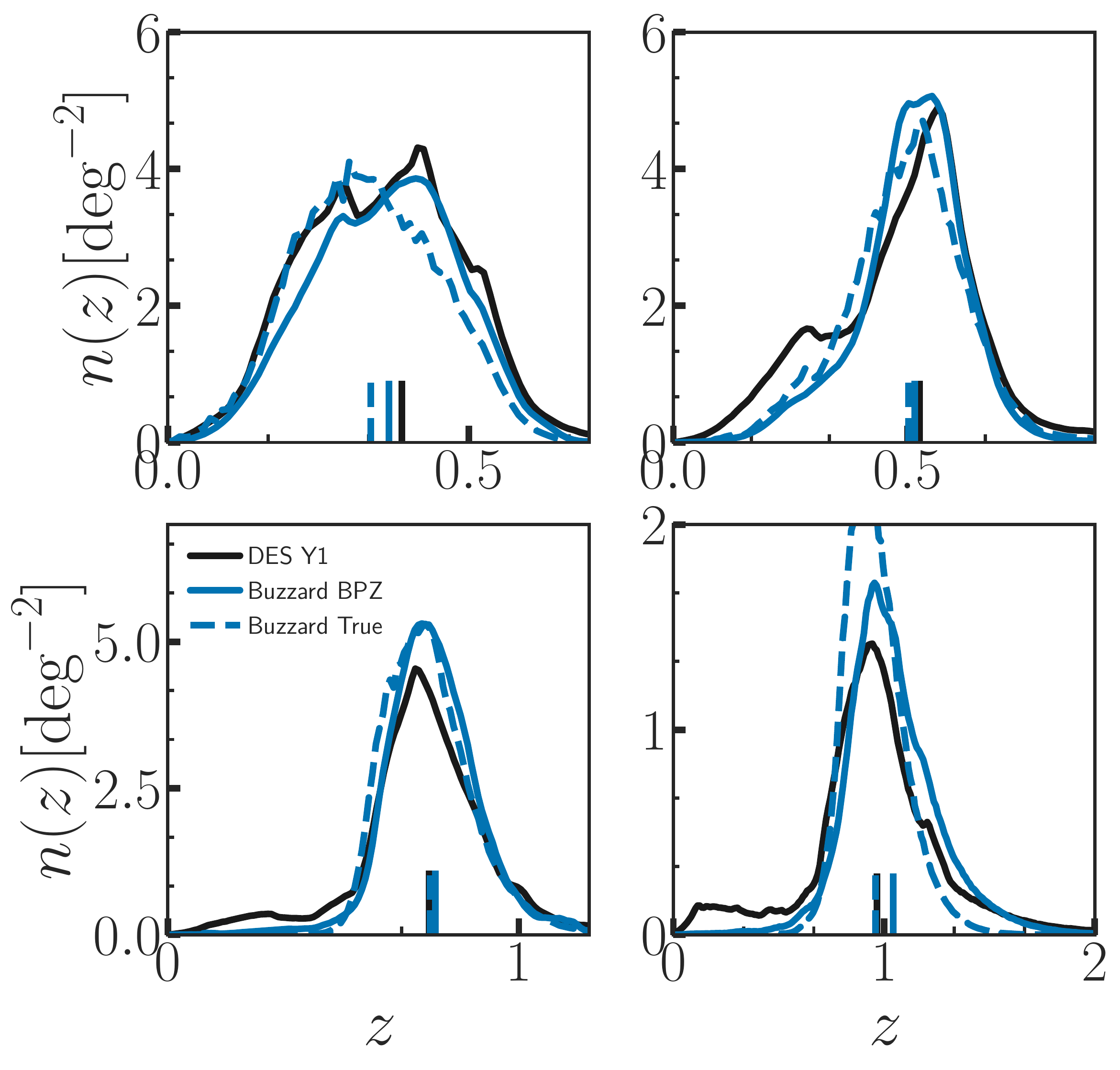}
  \caption{Photometric redshift performance.
  Left: Scatter and bias of BPZ photo-$z$ estimates as a function of redshift and
  magnitude for the simulated 'gold' galaxy sample. We do not attempt to compare
  this particular measurement to data, due to the lack of a complete spectroscopic
  validation set in the data. Right: The redshift distribution of the Buzzard \textsc{metacalibration} sample in the four source bins, ranging from low to high redshift
  from right to left and top to bottom, used for the Y1 $3\times2$ point analysis. These
  are estimated by stacking random draws from $p(z)$ as determined by BPZ (solid
  blue), and from true redshifts (dashed blue) compared to the $n(z)$ estimated
  for the DES Y1 \textsc{metacalibration} sample by stacking random draws from
  $p(z)$ as determined by BPZ (black). Vertical lines represent the means of each
  of these distributions.}
  \label{fig:pofz}
\end{figure*}

We calculate photometric redshifts for each object using BPZ
\citep{Benitez2000}, the primary photometric redshift code used in the DES Y1
analyses. The same BPZ configuration as used in the Y1 data is applied to the simulations. In particular
we use the same template SEDs as described in \citet{Hoyle17} and the bandpasses
and photometric calibration uncertainties described in \citet{y1gold}. In Fig.~
\ref{fig:pofz}, we show the characteristic error and bias on the \textsc{BPZ}
redshifts as a function of magnitude and $z_{\textsc{BPZ}}$ for all galaxies
detected in our DES Y1 catalogs after applying the error model described in
Appendix ~\ref{sec:photoerr}.

In order to facilitate analyses requiring a Y1-like source galaxy sample, we
have selected a sample of galaxies from our simulations matched to the DES Y1
\textsc{metacalibration} sample. The DES Y1 \textsc{metacalibration} sample is
the main shear catalog used in the DES Y1 $3\times2$ point analyses. This is a subsample
of the DES Y1 \textsc{gold} catalog \citep{y1gold} selected to have robust ellipticity 
measurements that can be used in the measurement of weak-lensing statistics. 
The shape measurement and subsequent selection algorithm is described in 
\citet{Zuntz2017}. As we do not perform image simulations using our simulated
catalogs, we cannot select a shear catalog in the same way as is done in the data.
Instead, we have chosen to select galaxies with similar signal-to-noise properties
as found in the data by performing the following cuts:

\begin{enumerate}
\item Mask all regions of the footprint where limiting magnitudes and PSF sizes cannot be estimated.
\item $\sigma(m_{{r,i,z}}) < 0.25$
\item $\sqrt{r_{gal}^{2} + (0.5~ r_{PSF})^{2}} > 0.75~r_{PSF}$
\item $m_r < 20.88 +  2.89~z$
\end{enumerate}

where $\sigma(m_{{r,i,z}})$ are the magnitude errors in $riz$ bands and $r_{PSF}$ is the $r$-band PSF FWHM estimated from the data at the position of each galaxy. $r_{gal}$ is the half light radius of the galaxy. The first three cuts are well motivated physically and intended to approximate signal-to-noise cuts imposed either explicitly or implicitly in shape catalog production on the data. These cuts yield a shape catalog that has too many galaxies compared to the data, possibly due to the fact that we have neglected to incorporate the dependence of photometric errors and detection on surface brightness into our photometric error model. The fourth cut is necessary to match the number density of sources in the four source redshift bins used in the DES Y1 key project, yielding values for the shot noise, $\sigma_{e}^2/n_{eff}$, in each bin of 0.050, 0.053, 0.047, 0.11, compared to 0.046, 0.059 0.046, and 0.11 for the \textsc{metacalibration} catalog on the Y1 data.

The extent to which this selection matches the DES Y1 \textsc{metacalibration}
redshift distributions can be seen in Fig.~\ref{fig:pofz}.
Qualitatively, the shape of the BPZ redshift distributions match those found in
the data, but there are some quantitative differences. The mean redshift in each source bin as estimated by BPZ on Buzzard is $0.368, ~0.515, ~0.762,\textrm{ and } 1.04$ while for the DES Y1 \textsc{metacalibration} sample they are $0.389, ~0.525, ~0.743$ and $0.966$. The differences in the mean redshifts of the four tomographic bins are statistically significant, with the errors on the mean in both the simulations and the data being on the order of $10^{-4}$. Having the true redshifts for every galaxy in our simulations, we can also compute the bias in this mean for each source bin, yielding offsets of 0.052, 0.023, -0.0037, 0.0061. These biases were also estimated in the data and found to be $-0.001\pm 0.016,~ -0.019 \pm 0.013, ~0.009\pm0.011,\textrm{ and } -0.018\pm-0.022$ \citep{Hoyle17, davis17b, gatti17}.

\begin{figure*}
  \includegraphics[width=\columnwidth]{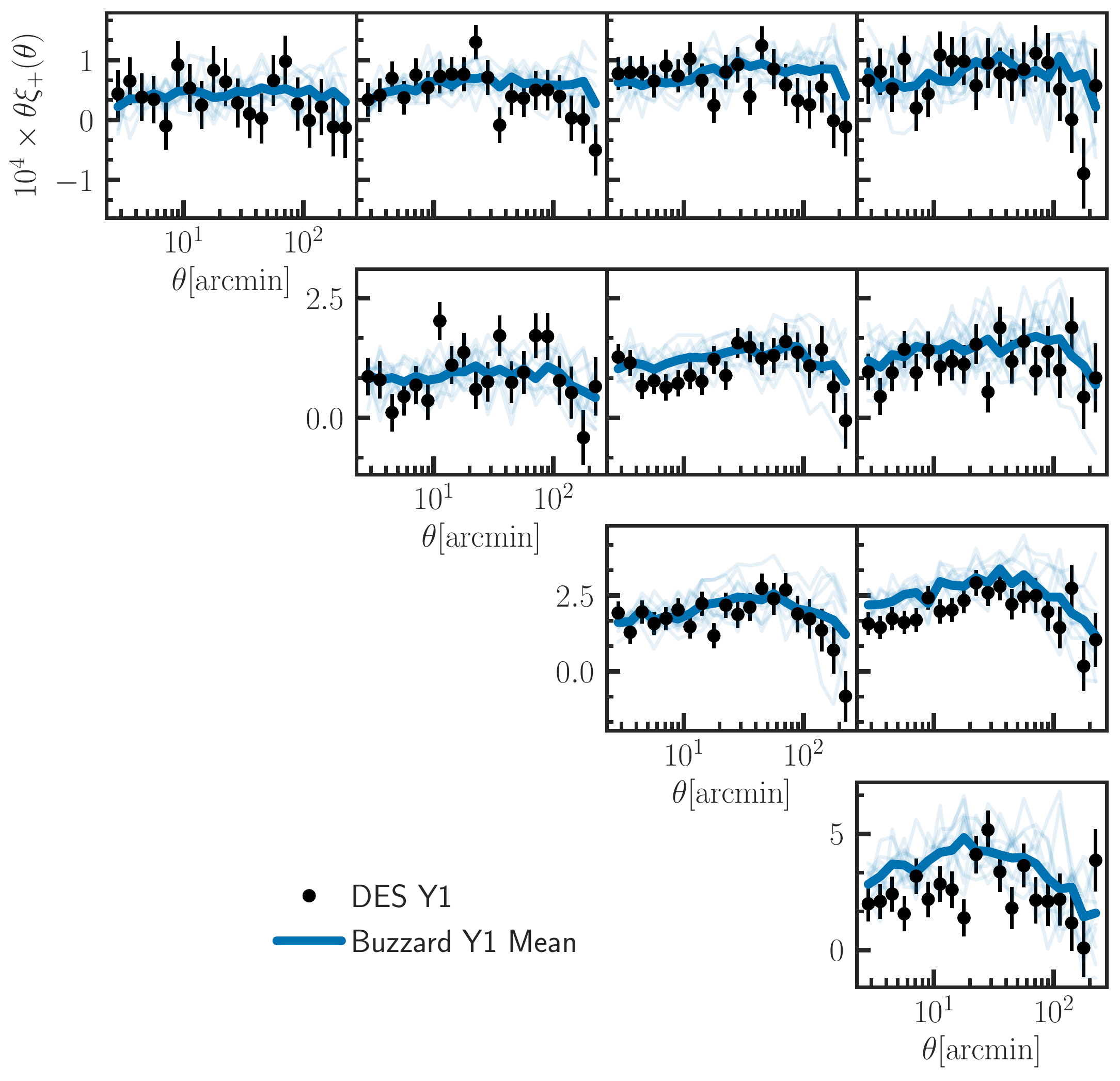}
  \includegraphics[width=\columnwidth]{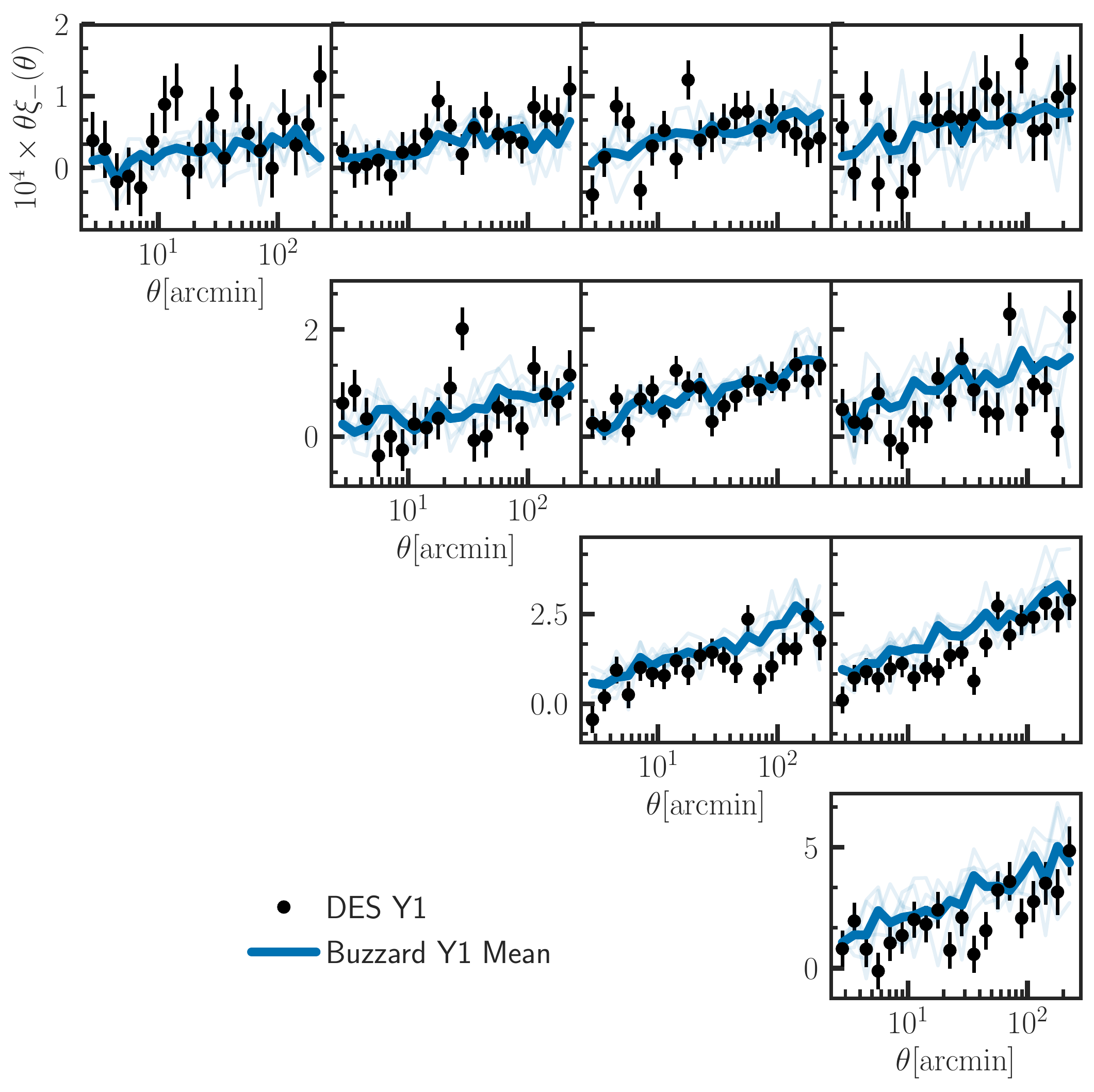}
  \caption{Comparison of $\xi_{+/-}$ tomographic auto- and cross correlations
    between the mean of all Buzzard simulations (solid blue) and DES Y1 (black).
    Light blue lines are measurements from individual simulations. The different 
    panels the unique cross-correlations between tomographic bins, where tomographic 
    bins go from low to high redshift from left to right and top to bottom.}
  \label{fig:xipm_metacal}
\end{figure*}

Comparisons of $\xi_{+/-}$ between Buzzard and the DES Y1 data can be seen in
Fig.~\ref{fig:xipm_metacal}. We refrain from making any statistical comparisons
between these sets of measurements as our intent was not necessarily to fit this
data, but note that in general the qualitative agreement between them is good. Differences in cosmology between our simulations and the Universe could contribute to the minor differences here, but coincidentally the Buzzard cosmology is quite close to the best fit cosmology as measured in \citet{y1kp}.

\subsection{redMaGiC}
\label{sec:redmagic}

\begin{figure}
  \includegraphics[width=\columnwidth]{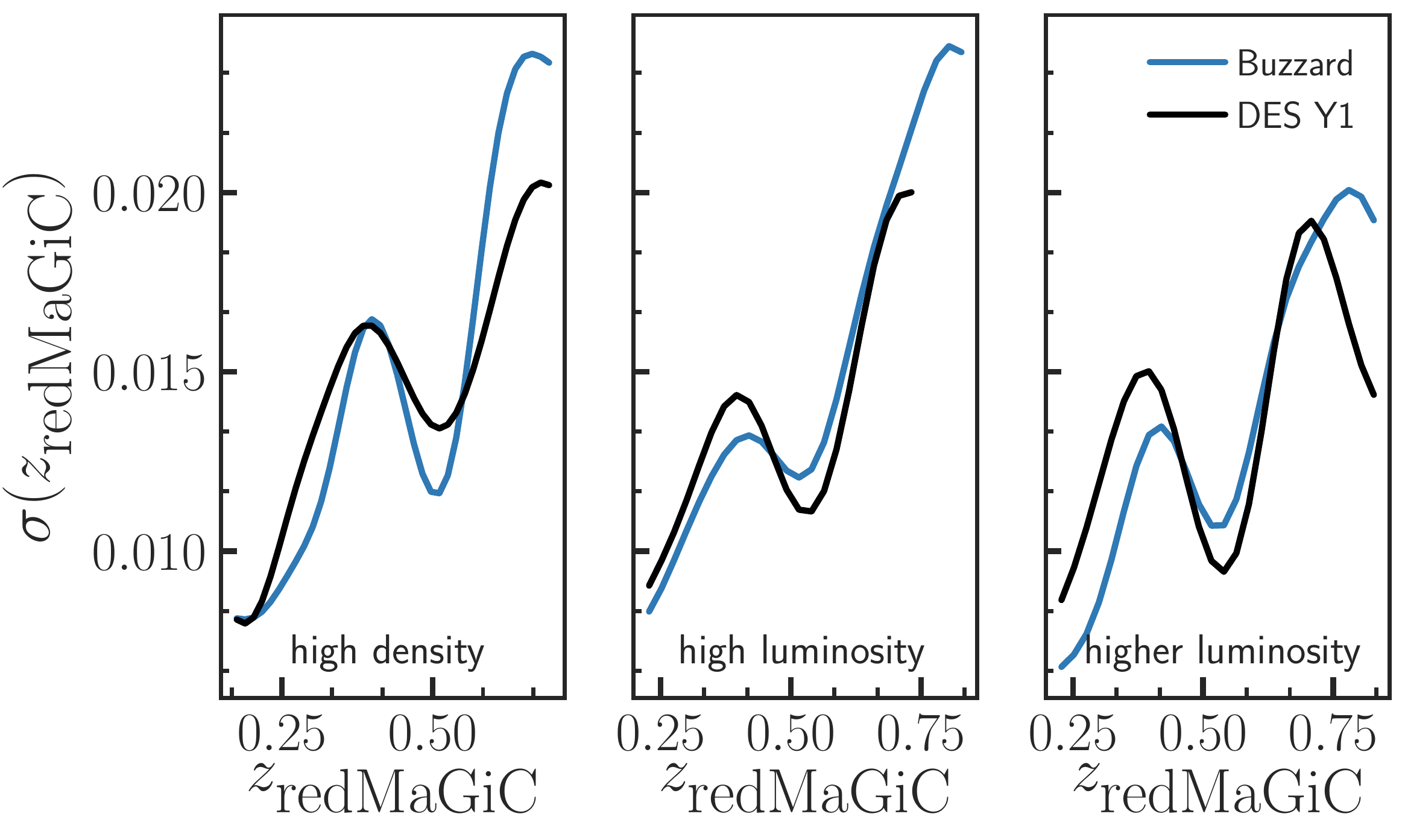}
  \caption{Comparison of \textsc{redMaGiC} photo-$z$s between Buzzard (blue) and DES Y1 (black). The three
    columns are the \textsc{redMaGiC} high density, high luminosity and higher luminosity
    samples from left to right. The figure shows the estimation of the photo-z errors as reported by the \textsc{redMaGiC} algorithm.}
  \label{fig:zredmagic}
\end{figure}

\begin{figure*}
  \includegraphics[width=\columnwidth]{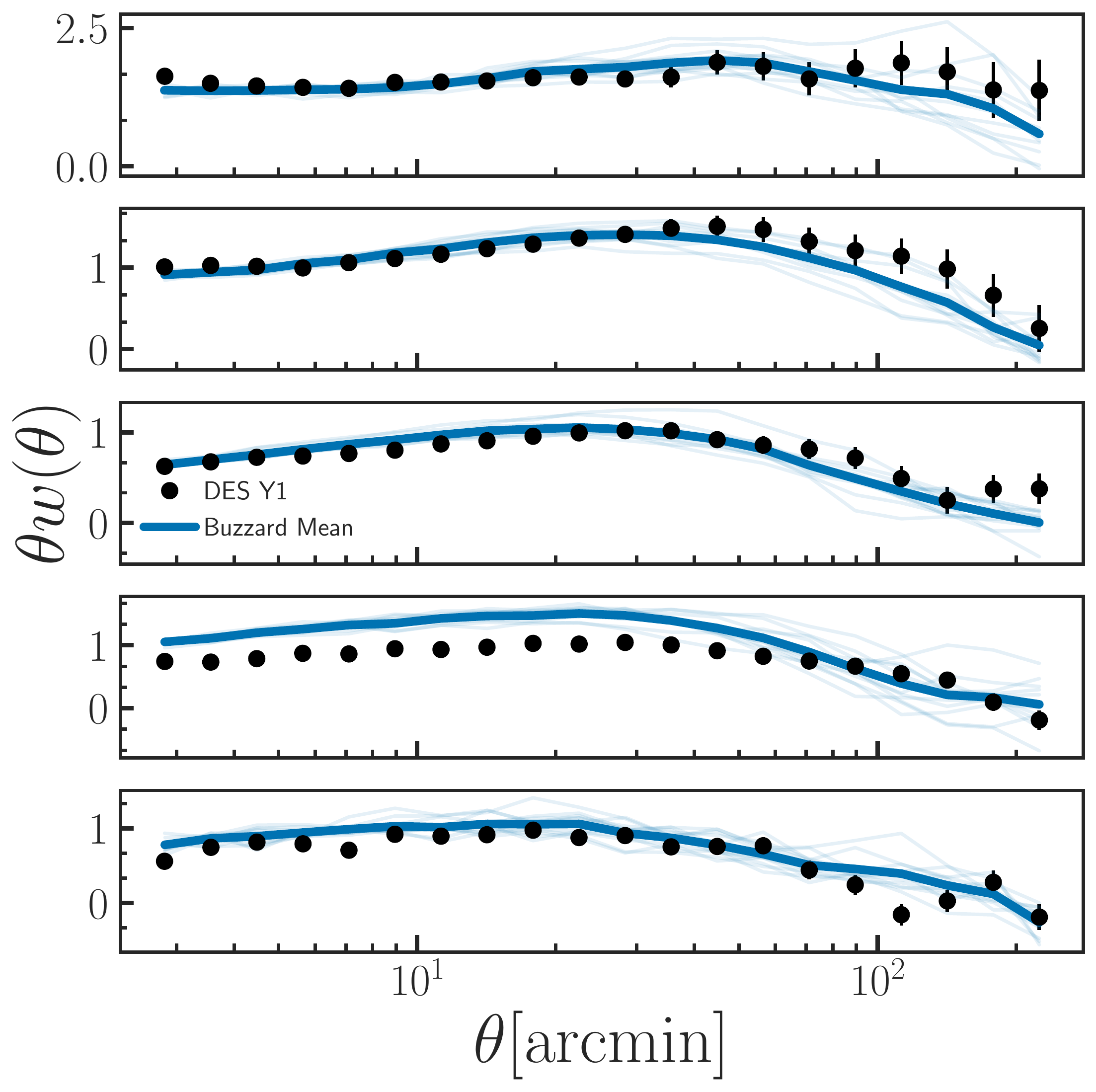}
  \includegraphics[width=\columnwidth]{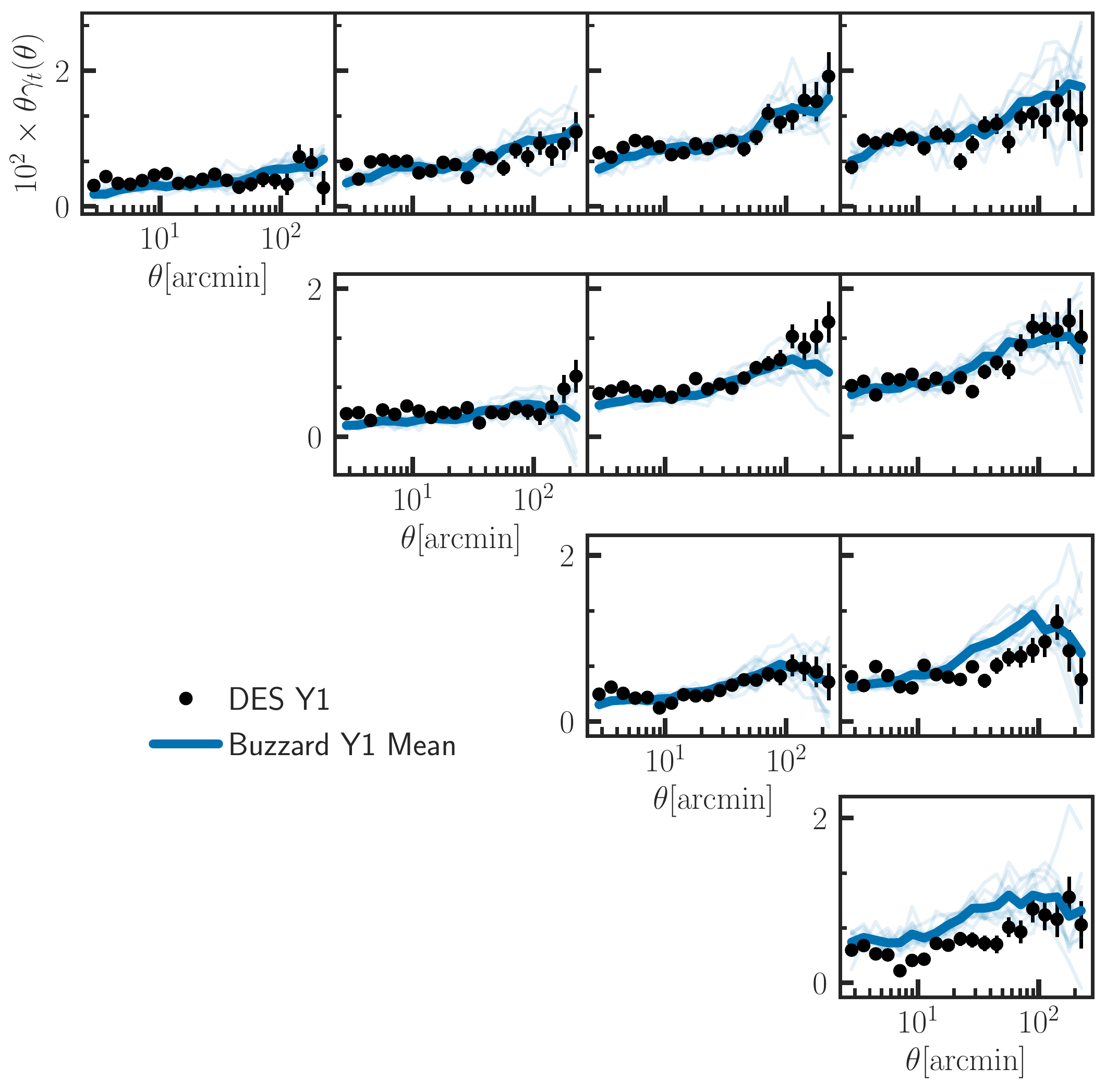}
  \caption{Left: Comparison of redMaGiC sample clustering between Buzzard and DES
    Y1. Top to bottom is low to high redshift with bins of 0.15-0.3, 0.3-0.45,
    0.45-0.6, 0.6-0.75, 0.75-0.9. Right: Comparison of galaxy--galaxy lensing around
    redMaGiC sample between buzzard and DES Y1 for all lens--source configurations
    with lenses in front of sources, using only the first 4 lens bins for
    readability.}
  \label{fig:wtheta_gammat_redmagic}
\end{figure*}

We also apply the \textsc{redMaGic} selection algorithm to our simulations. In
this case, unlike for the \textsc{metacalibration} sample, it is possible for us
to use the same selection algorithm in the simulations as is used in the data.

The \textsc{redMaGiC} galaxy sample was used extensively in the DES Y1 cosmology
analyses as a sample with robust photometric redshifts. As such we would
like to validate that our simulations reproduce this performance. As
discussed in Sec.~\ref{sec:redmapper}, the \addgals\ algorithm produces a red
sequence very similar to that found in the data, and thus the scatter of
the \textsc{redMaGiC} photo-$z$s is also similar as can be seen in
Fig.~\ref{fig:zredmagic}. Here we cite $\sigma(z_{redmagic})$ as the mean
of the internal \textsc{redMaGiC} estimate of this quantity in bins of $z_{redmagic}$. \textsc{redMaGiC} determines this quantity for each galaxy from a combination of the width of the red sequence, photometric errors on that galaxy, and an after-burner calibration on \textsc{redMaPPer} photometric redshifts. We find that $\sigma(z_{redmagic})$ in Buzzard is slighty smaller than its counterpart in the DES Y1 data at low redshift and slightly higher at high redshift. The level of discrepancy between the simulations and the data in this metric has yet to pose significant issues to DES analyses using these simulations. As measurements
become more precise, the underestimates of photo-$z$ errors here could lead to over-optimism in systematics estimation using Buzzard.

We also wish to compare our simulated \textsc{redMaGiC} sample to the data at
the level of the statistics used for the DES Y1 cosmology analyses, namely
angular clustering, $w(\theta)$, and galaxy--galaxy lensing,
$\gamma_{t}(\theta)$. We refer the reader interested in the details of these
measurements to \citet{MacCrann17} and \citet{y1kp}. In
Fig.~\ref{fig:wtheta_gammat_redmagic} we compare the mean value of $w(\theta)$
over our 18 simulated samples with the data in the five lens redshift bins that
were used in the DES Y1 clustering data vector. The first three bins use the
high density \textsc{redMaGic} sample, the fourth uses the high luminosity
\textsc{redMaGiC} sample and the higher luminosity sample is used in the fifth
bin. Because the \textsc{redMaGiC} photo-$z$ performance in the simulations
closely matches what is found in the data, differences in clustering here can be
interpreted as differences in the (non-linear) bias of our simulations and the real universe. In the first, second, third and fifth redshift bins the match between the simulations and the data is good. In the fourth bin, the simulations have an excess of clustering with respect to what is found in the data. While a large number of \textsc{redMaGiC} galaxies are placed in unresolved halos in our simulations, given that this sample is thought to populate halos down to masses of approximately $M_{\rm vir}\approx 10^{12}\hmsun$, it is unlikely that the discrepancies between our simulations and the data as seen here are due to resolution effects given the resilience of the \addgals\ method to resolution for large-scale clustering as demonstrated in \citet{wechsler_etal:19}. Instead, we observe that the \textsc{redMaGiC} sample in our simulations in the fourth bin is somewhat brighter than that observed in the data, giving a plausible explanation for the higher values of bias, given that brighter galaxies are in general more clustered.

Additional tests have shown that above $10\, \hmpc$
the bias of the \textsc{redMaGic} sample in the simulations conforms to a linear
bias model. Discrepancies such as those exhibited here may be relevant when 
estimating the impact of effects such as non-linear bias on various analyses,
thus these simulations should not be used to \textit{prove} that any given analysis
is immune to such systematics. Rather, the Buzzard simulations represent a plausible testing ground in which to perform necessary but not sufficient tests of the efficacy of an analysis. We also present measurements of $\gamma_{t}$ in Fig.~\ref{fig:wtheta_gammat_redmagic}, where we only show lens--source combinations with the sources behind the lenses, and we have not included the
5th lens bin for clarity of presentation. Again, the data and simulations are in good agreement.

These comparisons demonstrate that our simulated \textsc{redMaGiC} sample does
indeed resemble the sample as selected in the data. This is a non-trivial success of
these simulations as this level of agreement requires matches
between nearly every aspect of our simulations, including colors, luminosities,
and galaxy--halo connection, and the data. Although we have shown that there are
some quantitative discrepancies between the simulations and the data, these
tests show that this sample can be used as a realistic proxy for data when
developing and validating algorithms using the \textsc{redMaGiC} sample.

\subsection{redMaPPer}
\label{sec:redmapper}

\begin{figure}
  \includegraphics[width=\columnwidth]{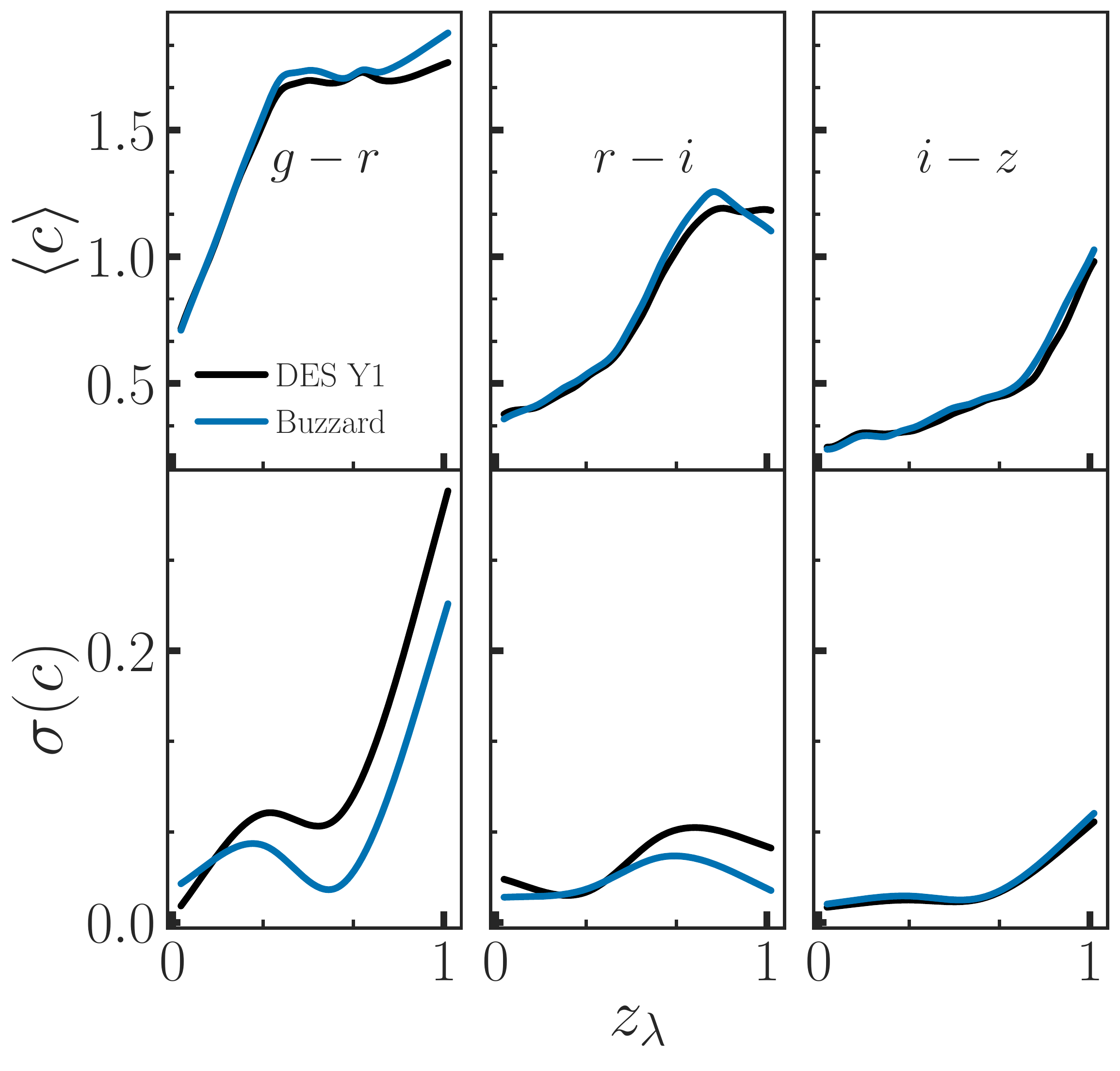}
  \caption{A comparison of the mean and scatter of red-sequence galaxy colors 
    between Buzzard and DES Y1 redMaPPer samples.}
  \label{fig:redsequence}
\end{figure}

\begin{figure*}
  \includegraphics[width=\columnwidth]{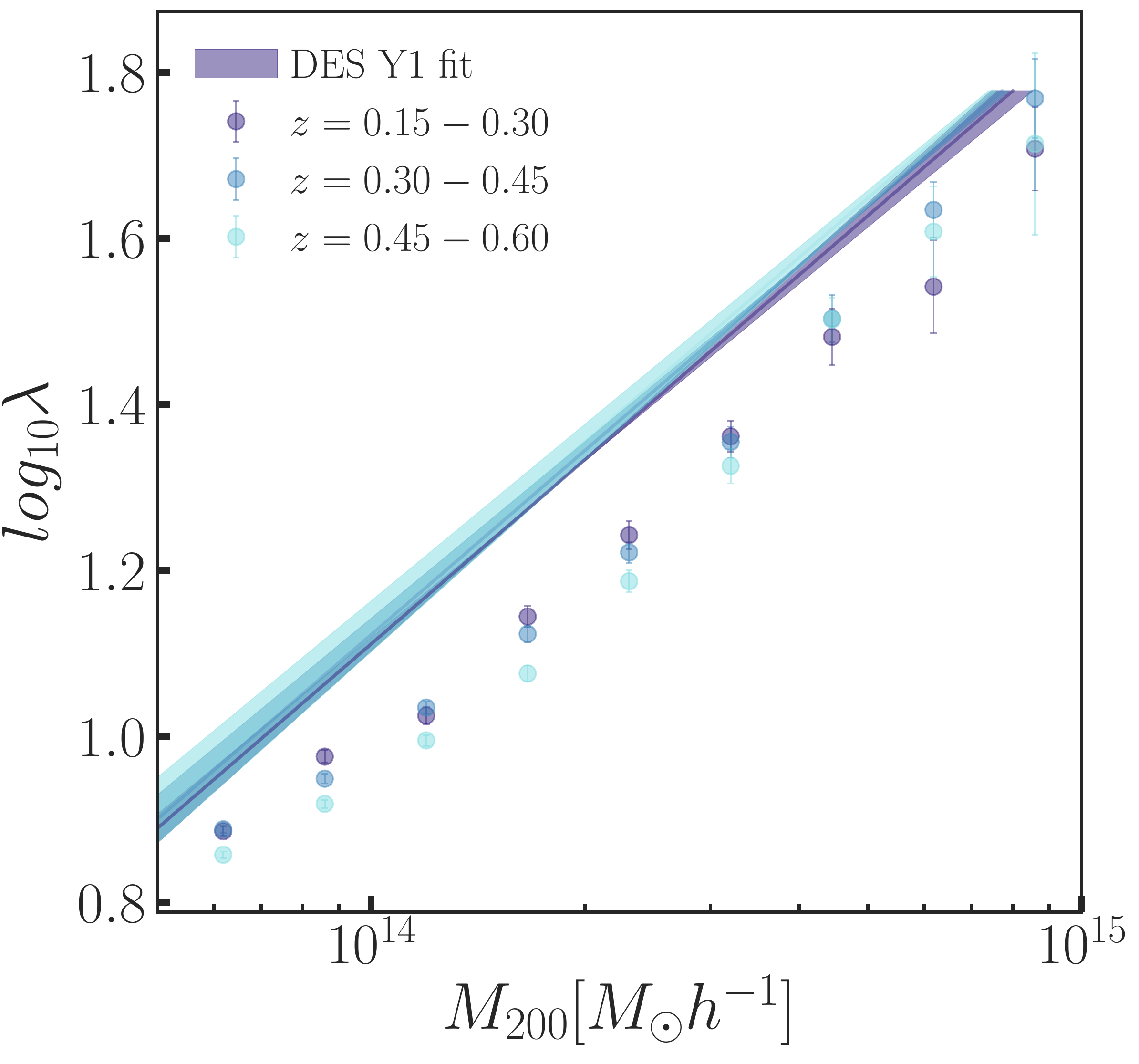}
  \includegraphics[width=\columnwidth]{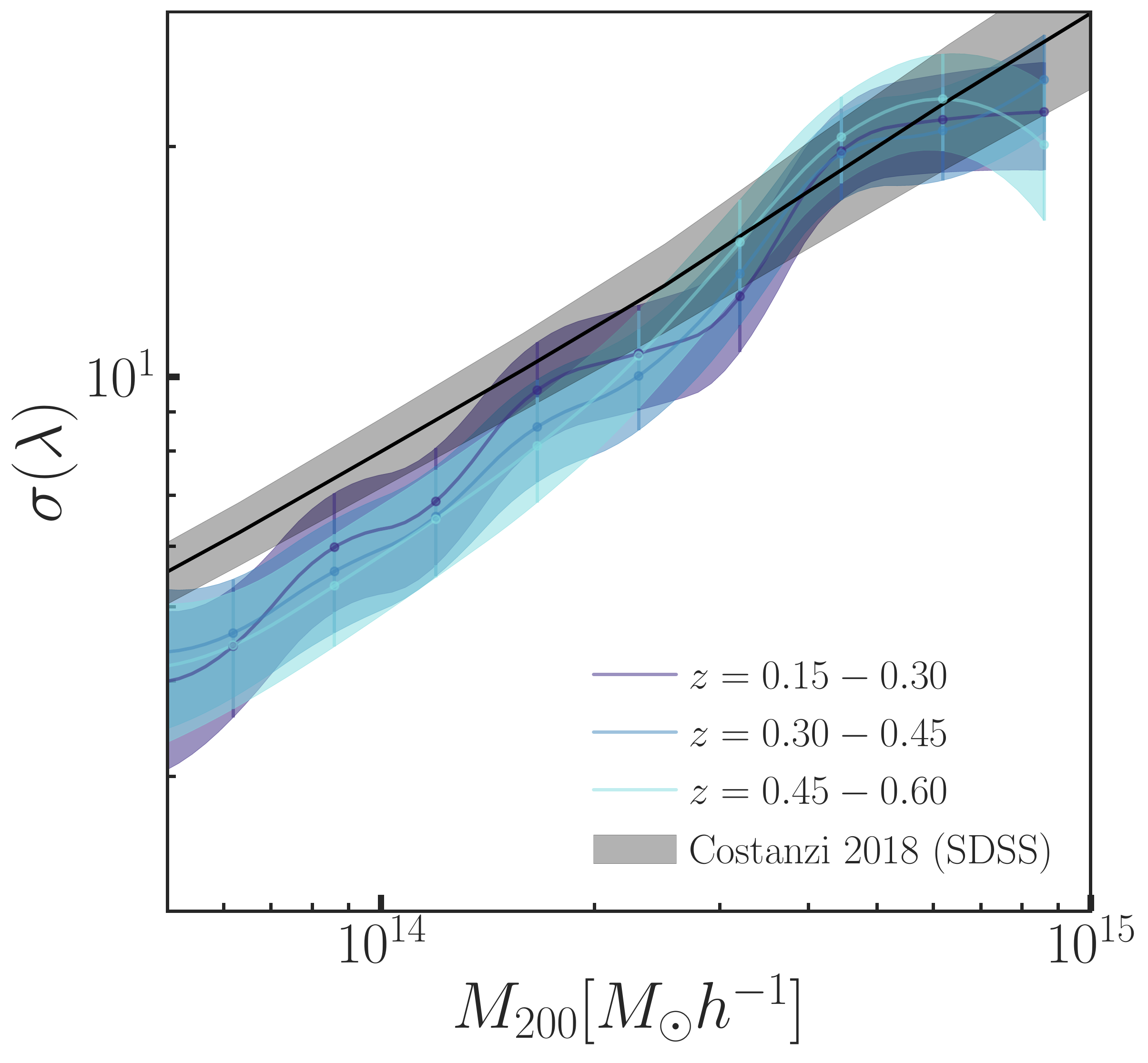}
  \caption{redMaPPer mass-$\lambda$ relationship.
    {\em Left:} The mass--richness relation in Buzzard compared to that inferred in
    DES Y1 redMaPPer clusters. The model fits to the data are the solid lines,
    the simulations are the points with error bars given by the error on the mean. 
    {\em Right:} Intrinsic scatter in the Buzzard mass--richness relation as a function of redshift compared to low redshift constraints from SDSS  
    \citep{Costanzi2018}. Lines between simulation measurements are quadratic interpolations.}
  \label{fig:massrichness}
\end{figure*}

\begin{figure*}
  \includegraphics[width=\columnwidth]{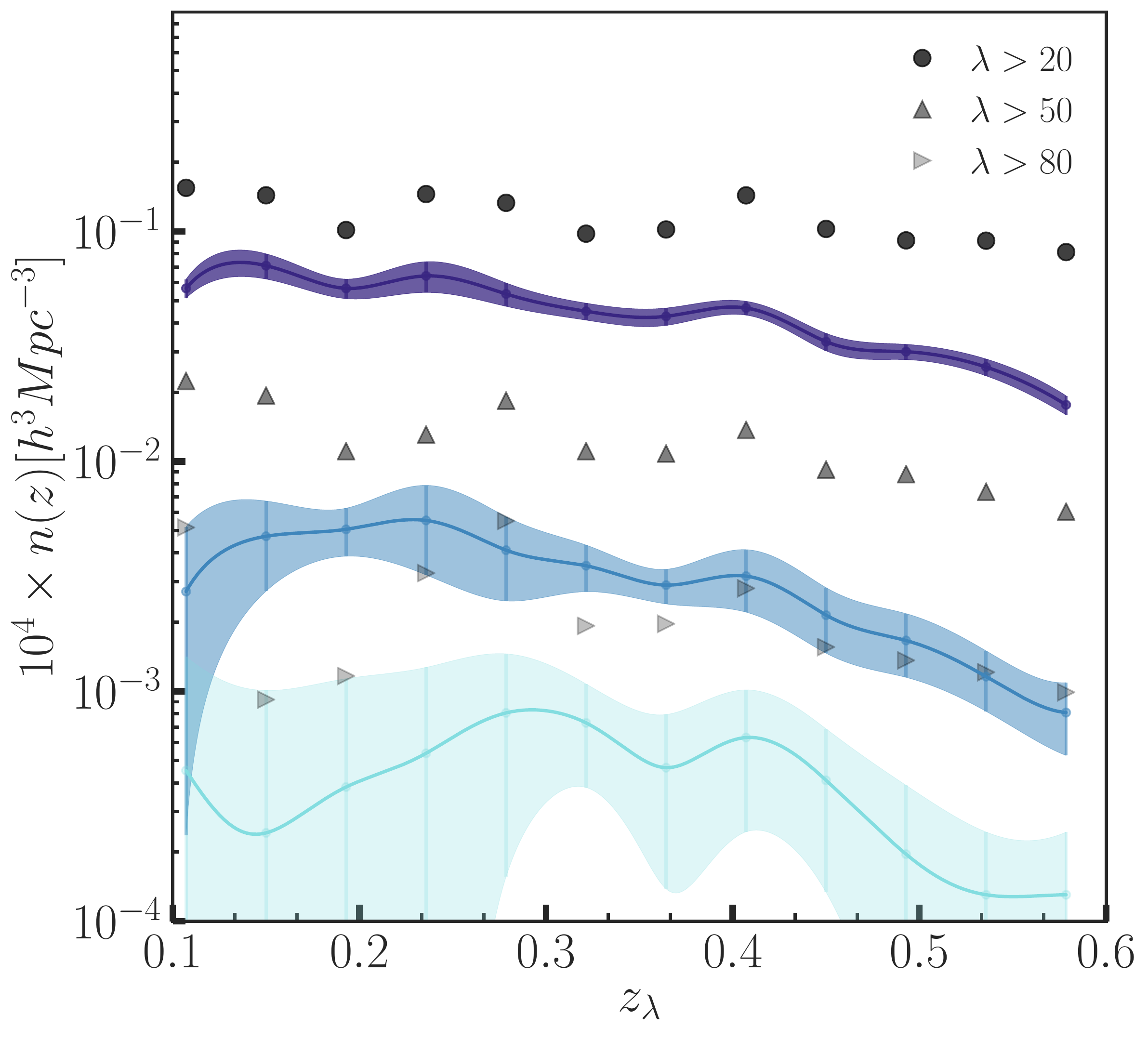}
  \includegraphics[width=\columnwidth]{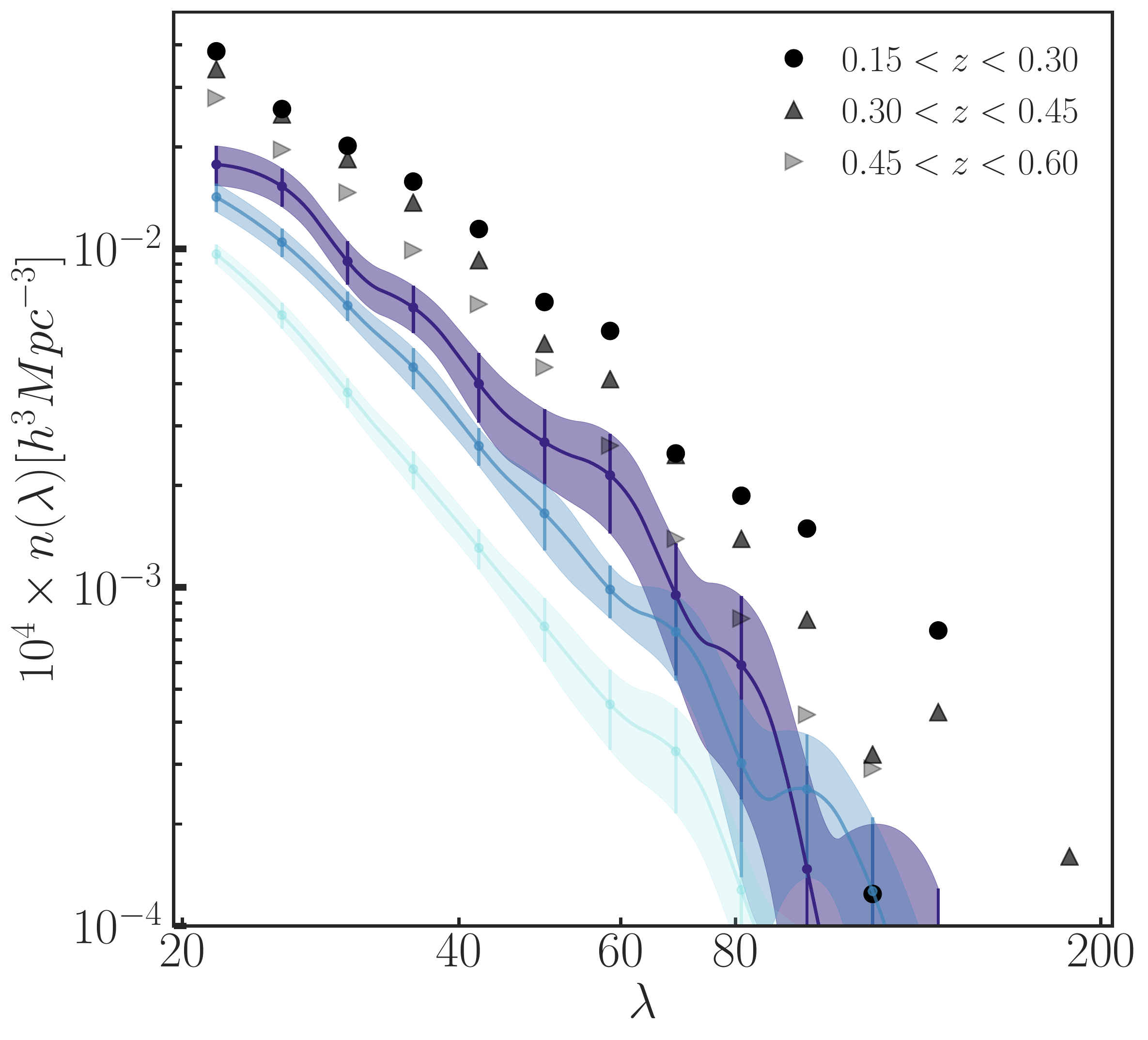}
  \caption{redMaPPer $\lambda$ abundance function.
    {\em Left:} The number of \textsc{redMaPPer} identified clusters for a few richness cuts as
    a function of redshift in Buzzard (lines) compared to the DES Y1 data (black points).
    Higher transparency points represent higher richness measurements.
    {\em Right:}
    Counts of the number of \textsc{redMaPPer} identified clusters in Buzzard (lines) 
    compared to the DES Y1 data (black points) as a function of richness in three different 
    redshift bins. Higher transparency points represent higher redshift measurements.}
  \label{fig:clustercounts}
\end{figure*}

The low computational cost of the pipeline outlined in this work also makes it
ideal for use in the study of photometrically identified cluster catalogs, where
very large-volume simulations are required to match the statistics in the data.
In order to facilitate these studies, we have run the \textsc{redMaPPer} cluster
finder on our simulations using the same configuration as used in the DES Y1
data. 

In order to produce simulated cluster catalogs whose properties match the
data, a number of observables must be reproduced at high fidelity. Firstly, a
tight red sequence must be present in the simulated galaxies' color distribution.
Our method for populating simulations with SEDs produces realistic color
distributions, especially for quenched galaxies whose rest-frame SEDs do not
seem to evolve much with redshift from their low redshift counterparts. This allows
us to reproduce the galaxy red sequence to high fidelity, as seen in
Fig.~\ref{fig:redsequence}. At high redshift the scatter in the red sequence of
the data appears larger than that observed in the simulations, especially in
$g-r$. One possible explanation for this is that our SEDs are constrained best
in the rest-frame optical, where the SEDs of the SDSS spectroscopic sample used
for our training set are measured. At high redshift, the $g$ and $r$ bands are
measuring the rest frame UV, and as such there is the possibility for
significant deviations of these bands from what is predicted from our training
set.

In addition to running \textsc{redMaPPer} with the standard configuration used
on the data, we have also produced a catalog that uses the same red sequence
model, but measures richnesses around the true halo centers in our simulations.
By doing this we can learn about the $\lambda - M_{200b}$ relationship without
the complications imparted by mis-centering due to structures projected along 
the line-of-sight. In Fig.~\ref{fig:massrichness}, we present a comparison of the
measured $\lambda - M_{200b}$ relation using true halo centers with that inferred
using the DES Y1 data \citep{McClintock2018}. We see a deficit in richness at fixed mass when compared to the DES Y1 measurements. This is due to the inability of our model to perfectly reproduce quenched galaxy clustering, leading to fewer red satellite galaxies in massive halos than found in the data \citep{wechsler_etal:19}.  

In the right hand panel, we show measurements of the scatter in the 
$\lambda_{obs}-M_{200b}$ relation in our simulations.
The DES weak-lensing cluster mass calibration presented in \citet{McClintock2018}
does not constrain the scatter in this relation, so we cannot compare explicitly 
to DES data, but these values for scatter are consistent with those presented in 
other \textsc{redMaPPer} analyses, such as \citet{Costanzi2018} (shown in black), showing a marginal preference for lower scatter at low halo masses than what is found in the data. 

In Fig.~\ref{fig:clustercounts} we present a comparison between the mean 
number densities of the 18 simulated \textrm{redMaPPer} catalogs and the data as
a function of redshift and cluster richness, $\lambda$. For these comparisons,
we use \textsc{redMaPPer} catalogs run with the same configuration used on the data.
The simulations under-predict the data number densities with the discrepancy becoming worse at high redshift. This is consistent with the under-prediction of richness at fixed halo mass as observed in Fig. \ref{fig:massrichness}. This is likely additionally
exacerbated by marginally lower scatter in richness at fixed mass in our simulations compared to the data. Aside from the observed discrepancy between number densities of clusters in the data and our simulations, the redshift evolution of these quantities is in the same direction, with the simulations showing more redshift evolution at high redshift than the data.

Overall, the simulations presented here have more difficulty matching
\textsc{redMaPPer} observables compared to the other samples presented in this
work due to their sensitivity to the spatial dependence of galaxy colors down to small
scales. Future improvements to these simulations aimed at improving these
observables will focus on new methods for assigning galaxy SEDs to their correct
locations in the cosmic web and for using larger samples of data to constrain
one-point statistics such the luminosity functions and $f_{red}$ so that rare
galaxy populations such as cluster members are sampled with higher statistics. 
Nonetheless, these simulations represent the state of the art in reproducing the
cluster observables shown here. We are currently using these simulations to test
aspects of the DES cluster cosmology analysis. These simulations should not be 
used to calibrate analyses, but the qualitative agreement between the observables 
presented here demonstrates that they can be used as a plausible simulated universe
on which to test analysis pipelines and develop algorithms.

\section{Summary}
\label{sec:summary}

This paper presents a suite of 18 synthetic DES Y1 catalogs out to $z = 2.35$ and
to a depth of $r \sim 26$ (excluding very bright objects at $z > 2.35$). They include:
\begin{itemize}
\item halo and particle catalogs
\item galaxy positions, magnitudes, ellipticities, and sizes
\item lensing for each galaxy via a curved sky ray-tracing algorithm
\item realistic photometric errors, masking and photometric redshifts
\item \textsc{redMaGiC}, \textsc{redMaPPer}, and weak-lensing source samples
\end{itemize}
We have demonstrated the fidelity of these catalogs by comparing relevant observables to data or theory where applicable, with a focus on those tests that are most relevant for the cosmological analysis of current DES data:
\begin{itemize}
\item matter power spectra, halo mass functions and halo bias
\item galaxy magnitude and color distributions
\item photometric redshift distributions
\item $3\times2$ point observables in the DES Y1 lens and source bins
\item \textsc{redMaPPer} mass-observable relation and number densities
\end{itemize}

This provides a high-fidelity reproduction of the DES Y1 data that facilitates
study of many large-scale structure probes simultaneously, including galaxy
clustering, optically selected galaxy clusters, shear correlation functions, and the lensing profiles of galaxies and clusters.

The limiting factors in the fidelity of these simulations are twofold. The first limitation is our methodology for tuning the free parameters of our galaxy model. Ideally, one would optimize these parameters jointly by tuning to the observations
that we wish to reproduce, e.g. DES 3x2-point observables and cluster abundances, 
via an iterative optimization scheme. 
This is currently infeasible, as the pipeline beginning with
our galaxy model and including necessary observational effects such as
ray-tracing, photometric errors, photometric redshift estimation and sample
selection, takes much too long to run such an optimization algorithm on. Instead
we have settled for tuning these parameters individually, sometimes by hand, e.g.
in the case of the red-fraction of galaxies. 
Progress in optimizing the parameters of \addgals\ may either be made by
constructing fast proxies for the relevant observables as a function of our
model parameters, or by simply speeding up the process of running each simulation.

The second limitation is the simplifying assumptions that we have made to extrapolate
our model to high redshift. This is manifest in a few ways, in particular 
the discrepancies between the colors in Buzzard and COSMOS at high redshift as
detailed in Sec. \ref{sec:colors}, and discrepancies with the data in observed
magnitude distributions (Sec. \ref{sec:lf}) and clustering (Sec.
\ref{sec:redmagic}). Progress here will necessitate updates to galaxy models, in
particular the SEDs that we use at high redshift and our parameterization of the
galaxy--halo connection. These upgrades are already underway.

The strategy we pursued here was designed to work with moderate resolution
simulations that can be run on relatively standard clusters today. The rough
computation time from initial conditions to a final catalog is $\sim$ 150,000
CPU hours, and we can complete this end-to-end pipeline in approximately one
week.  It has already allowed the efficient production of many times the DES
volume, which has proven essential for developing realistic error estimates from
surveys. This strategy will also allow us to repeat the catalog creation a
number of times with differing choices of the cosmological model, the galaxy
evolution model, and the model for inclusion of systematic errors.

There are however significant scientific motivations for higher resolution
simulations, which allow for more accurate models of the galaxy--halo
connection. Ideally, one would model the entire galaxy population presented here
in simulations that resolve galactic substructures (for example, see the low
redshift synthetic galaxy distribution presented in \citealt{Reddick12}). In
\cite{wechsler_etal:19}, we have reviewed the computing challenges that must be overcome to achieve this and the limitations of the specific resolution choice made here. We emphasize that the pipeline we have developed here is modular, and can be
readily extended, including to higher resolution simulations.

Upon posting of this article we are making the simulations described here available upon request. This includes the underlying $N$-body simulations, the 10,313 square degree galaxy catalogs, as well as the \textsc{metacalibration}, \textsc{redMaGiC}, and \textsc{redMaPPer} catalogs, random points, and the $3\times2$ point data vectors presented in this paper.

We will make the aforementioned data products freely downloadable at \url{BuzzardFlock.github.io} at the 
time this study and its companion papers are published\footnote{Interested users of 
the catalog before this time are encouraged to contact the authors at
\href{mailto:jderose@stanford.edu}{jderose@stanford.edu} or
\href{mailto:rwechsler@stanford.edu}{rwechsler@stanford.edu}}. We expect that
this can be of use in a wide range of studies of large-scale structure, galaxy
clusters, weak lensing, and photometric redshift distributions.

\acknowledgements

We are very grateful to numerous Dark Energy Survey collaborators who provided useful tests of these catalogs during various stages of development.  These include Joanne Cohn, Chihway Chang, Joerg Dietrich, Tim Eifler, Oliver Friedrich, Daniel Gruen, Jiangang Hao, Elisabeth Krause, Boris Leistedt, Eduardo Rozo, Eusebio Sanchez, Michael Troxel, Martin White, and Yuanyuan Zhang.

Some of the computing for this project was performed on the Sherlock 
cluster. We would like to thank Stanford University and the Stanford
Research Computing Center for providing computational resources 
and support that contributed to these research results. This 
research used resources of the National Energy Research Scientific
Computing Center (NERSC), a U.S. Department of Energy Office of
Science User Facility operated under Contract No. DE-AC02-05CH11231.
Some of the cosmological simulations were also run on the XSEDE machine
Ranger. This work was completed in part with resources provided by the 
University of Chicago Research Computing Center. We are grateful to 
Stuart Marshall and the rest of the SLAC
computing team for extensive support of this work. Application support
was provided by the Advanced Support for TeraGrid Applications.  
We thank Volker Springel for providing the L-Gadget2 code.

JD, RHW, MB, MRB received support from the U.S. Department of Energy under
contract number DE-AC02-76SF00515, including support from a SLAC LDRD
grant, and from a Terman Fellowship at Stanford University.  MECS was
supported by the NSF under Award No. AST-0901965. Argonne National 
Laboratory's work was supported by the U.S. Department of Energy, 
Office of Science, Office of Nuclear Physics, under contract 
DE-AC02-06CH11357.

This study made use of the SDSS DR7 Archive, for which funding has
been provided by the Alfred P. Sloan Foundation, the Participating
Institutions, the National Aeronautics and Space Administration, the
National Science Foundation, the U.S. Department of Energy, the
Japanese Monbukagakusho, and the Max Planck Society. The SDSS Web site
is http://www.sdss.org/.  The SDSS is managed by the Astrophysical
Research Consortium (ARC) for the Participating Institutions: the
University of Chicago, Fermilab, the Institute for Advanced Study, the
Japan Participation Group, the Johns Hopkins University, Los Alamos
National Laboratory, the Max-Planck-Institute for Astronomy (MPIA),
the Max-Planck-Institute for Astrophysics (MPA), New Mexico State
University, University of Pittsburgh, Princeton University, the United
States Naval Observatory, and the University of Washington.

Funding for the DES Projects has been provided by the U.S. Department of Energy, the U.S. National Science Foundation, the Ministry of Science and Education of Spain, 
the Science and Technology Facilities Council of the United Kingdom, the Higher Education Funding Council for England, the National Center for Supercomputing 
Applications at the University of Illinois at Urbana-Champaign, the Kavli Institute of Cosmological Physics at the University of Chicago, 
the Center for Cosmology and Astro-Particle Physics at the Ohio State University,
the Mitchell Institute for Fundamental Physics and Astronomy at Texas A\&M University, Financiadora de Estudos e Projetos, 
Funda{\c c}{\~a}o Carlos Chagas Filho de Amparo {\`a} Pesquisa do Estado do Rio de Janeiro, Conselho Nacional de Desenvolvimento Cient{\'i}fico e Tecnol{\'o}gico and 
the Minist{\'e}rio da Ci{\^e}ncia, Tecnologia e Inova{\c c}{\~a}o, the Deutsche Forschungsgemeinschaft and the Collaborating Institutions in the Dark Energy Survey. 

The Collaborating Institutions are Argonne National Laboratory, the University of California at Santa Cruz, the University of Cambridge, Centro de Investigaciones Energ{\'e}ticas, 
Medioambientales y Tecnol{\'o}gicas-Madrid, the University of Chicago, University College London, the DES-Brazil Consortium, the University of Edinburgh, 
the Eidgen{\"o}ssische Technische Hochschule (ETH) Z{\"u}rich, 
Fermi National Accelerator Laboratory, the University of Illinois at Urbana-Champaign, the Institut de Ci{\`e}ncies de l'Espai (IEEC/CSIC), 
the Institut de F{\'i}sica d'Altes Energies, Lawrence Berkeley National Laboratory, the Ludwig-Maximilians Universit{\"a}t M{\"u}nchen and the associated Excellence Cluster Universe, 
the University of Michigan, the National Optical Astronomy Observatory, the University of Nottingham, The Ohio State University, the University of Pennsylvania, the University of Portsmouth, 
SLAC National Accelerator Laboratory, Stanford University, the University of Sussex, Texas A\&M University, and the OzDES Membership Consortium.

Based in part on observations at Cerro Tololo Inter-American Observatory, National Optical Astronomy Observatory, which is operated by the Association of 
Universities for Research in Astronomy (AURA) under a cooperative agreement with the National Science Foundation.

The DES data management system is supported by the National Science Foundation under Grant Numbers AST-1138766 and AST-1536171.
The DES participants from Spanish institutions are partially supported by MINECO under grants AYA2015-71825, ESP2015-66861, FPA2015-68048, SEV-2016-0588, SEV-2016-0597, and MDM-2015-0509, 
some of which include ERDF funds from the European Union. IFAE is partially funded by the CERCA program of the Generalitat de Catalunya.
Research leading to these results has received funding from the European Research
Council under the European Union's Seventh Framework Program (FP7/2007-2013) including ERC grant agreements 240672, 291329, and 306478.
We  acknowledge support from the Australian Research Council Centre of Excellence for All-sky Astrophysics (CAASTRO), through project number CE110001020, and the Brazilian Instituto Nacional de Ci\^encia
e Tecnologia (INCT) e-Universe (CNPq grant 465376/2014-2).

This manuscript has been authored by Fermi Research Alliance, LLC under Contract No. DE-AC02-07CH11359 with the U.S. Department of Energy, Office of Science, Office of High Energy Physics. The United States Government retains and the publisher, by accepting the article for publication, acknowledges that the United States Government retains a non-exclusive, paid-up, irrevocable, world-wide license to publish or reproduce the published form of this manuscript, or allow others to do so, for United States Government purposes.
\appendix

\section{Modeled Galaxy Properties in the Final Catalog}
\label{app:tags}

A list of modeled intrinsic and observed galaxy properties as well as halo properties included in the data release are provided in Tables 2-4. Parameters related to halo ellipticities and angular momenta are determined using all particles within $R_{\textrm{vir}}$. For more information about the released data including file formats and data access instructions, please see \url{BuzzardFlock.github.io}.

\begin{table}[h]
\label{tab:intrinsic-props}
\centering
  \begin{tabular}{ c | c | c }
    \hline
    Name & Description & Unit \\ \hline \hline 
   {\tt ID} & A unique identification number for each object. & N/A \\ \hline
   {\tt TRA} & Simulated right ascension (unlensed). & degrees \\ \hline
   {\tt TDEC} & Simulated declination (unlensed). & degrees \\ \hline
   {\tt Z\_COS} &  Redshift (only including Hubble flow) & N/A \\ \hline
   {\tt AMAG} & Galaxy absolute magnitude in DES grizY bands in z=0.1 frame & $h^{-1}$mag \\ \hline 
   {\tt TMAG} & True apparent galaxy magnitudes in grizY bands & mag \\ \hline
   {\tt MAG\_R} & SDSS $r$-band absolute magnitude in z=0.1 frame. & $h^{-1}$mag \\ \hline 
   {\tt P[XYZ]} & 3D comoving galaxy positions & $\hmpc$ \\ \hline
   {\tt V[XYZ]} & 3D physical velocity & $s^{-1}$km\\ \hline 
   {\tt TE} & Ellipticity in local {\tt RA} and {\tt DEC} directions & N/A \\ \hline 
   {\tt TSIZE} & Galaxy half-light radius & arcsec \\ \hline
   {\tt GAMMA1} & Weak lensing shear in the local \textsc{RA} direction & N/A \\ \hline
   {\tt GAMMA2} & Weak lensing shear in the local \textsc{DEC} direction & N/A \\ \hline
   {\tt KAPPA} & Weak lensing convergence & N/A \\ \hline 
   {\tt MU} & Weak lensing magnification & N/A \\ \hline 
   {\tt SEDID} & Index of SDSS galaxy from SED training set & N/A \\ \hline 
   {\tt COEFFS} & Coefficients of SED template from \textsc{kcorrect} & N/A \\ \hline 
   {\tt HALOID} & {\tt ID} of halo nearest to the galaxy in 3D & N/A \\ \hline 
   {\tt M200} & Halo mass (\textsc{M200b}) of the halo nearest to the galaxy in 3D & $\hmsun$ \\ \hline 
   {\tt R200}& Comoving radius of the halo nearest to the galaxy in 3D corresponding to \textsc{M200b} & $\hmpc$\\ \hline
   {\tt RHALO} & Distance to the nearest halo in 3D $\hmpc$ & $\hmpc$ \\ \hline 
   {\tt CENTRAL} & A flag indicating if the galaxy is the central
  galaxy of a resolved halo. & N/A \\ \hline \hline 
  \end{tabular}
  \caption{Intrinsic Galaxy Properties}
\end{table}

\begin{table}[h]
\label{tab:observed-props}
\centering
  \begin{tabular}{ c | c | c }
    \hline
    Name & Description & Unit \\ \hline \hline 
   {\tt Z} & Redshift (including redshift space distortions) & N/A \\ \hline 
   {\tt RA} & Lensed right ascension & N/A \\ \hline 
   {\tt DEC} & Lensed declination & N/A \\ \hline 
   {\tt LMAG} & Lensed apparent galaxy magnitude in DES grizY bands & mag \\ \hline 
   {\tt MAG\_[GRIZ]} & Noisy lensed apparent galaxy magnitude in DES grizY bands & mag \\ \hline 
   {\tt MAGERR\_[GRIZ]} & Galaxy magnitude errors in DES grizY bands & mag \\ \hline 
   
   {\tt FLUX\_[GRIZ]} & Noisy lensed galaxy fluxes in DES grizY bands & nanomaggies \\ \hline 
   {\tt IVAR\_[GRIZ]} & Inverse variance of galaxy fluxes in DES grizY bands & $\textrm{nanomaggies}^{-2}$ \\ \hline 
   {\tt EPISLON} & Lensed ellipticity in local {\tt RA} and {\tt DEC} directions & N/A \\ \hline
   {\tt SIZE} & Lensed galaxy half-light radius & arcmin \\ \hline 
   {\tt Z\_MEAN} & Mean redshift estimate as determined by BPZ & N/A \\ \hline 
   {\tt Z\_MC} & Monte Carlo draw from BPZ p(z) & N/A \\ \hline \hline 
   
  \end{tabular}
  \caption{Observed Galaxy Properties}
\end{table}

\begin{table}[h]
\label{tab:halo-props}

\centering
  \begin{tabular}{ c | c | c }
    \hline
    Name & Description & Unit \\ \hline \hline 
   {\tt HALOID} & A unique identification number for each halo & N/A \\ \hline
   {\tt RA} & Lensed right ascension & degrees \\ \hline 
   {\tt DEC} & Lensed declination & degrees \\ \hline    
   {\tt TRA} & Unlensed right ascension & degrees \\ \hline 
   {\tt TDEC} & Unlensed declination & degrees \\ \hline       
   {\tt Z} & Redshift including redshift space distortions & N/A \\ \hline 
   {\tt Z\_COS} & Redshift including only Hubble flow & N/A \\ \hline 
   {\tt M200} & Halo mass, $M_{200,\textrm{background}}$ & $\hinv\msun$ \\ \hline 
   {\tt R200} & Comoving halo radius, $R_{200,\textrm{background}}$ & $\hmpc$ \\ \hline 
   {\tt M200C} & Halo mass, $M_{200,\textrm{crit}}$ & $\hinv\msun$ \\ \hline 
   {\tt R200C} & Comoving halo radius, $R_{200,\textrm{crit}}$ & $\hmpc$ \\ \hline 
   {\tt M500C} & Halo mass, $M_{500,\textrm{crit}}$ & $\hinv\msun$ \\ \hline 
   {\tt R500C} & Comoving halo radius, $R_{500,\textrm{crit}}$ & $\hmpc$ \\ \hline 
   {\tt MVIR} & Halo mass, $M_{\textrm{vir}}$ & $\hinv\msun$ \\ \hline 
   {\tt RVIR} & Comoving halo radius, $R_{\textrm{vir}}$ & $\hmpc$ \\ \hline 
   {\tt M2500} & Halo mass, $M_{2500,\textrm{crit}}$ & $\hinv\msun$ \\ \hline 
   {\tt R2500} & Comoving halo radius, $R_{2500,\textrm{crit}}$ & $\hmpc$ \\ \hline 
   {\tt VRMS} & 3D velocity dispersion of particles in the halo within $R_{\textrm{vir}}$ & $s^{-1}$km \\ \hline 
   {\tt RS} & Comoving halo scale radius from NFW profile fit & $\hkpc$ \\ \hline 
   {\tt J[X,Y,Z]} & Halo angular momentum & $(\hinv\msun)(\hMpc)$ $s^{-1}$km \\ \hline 
   {\tt LAMBDA} & Halo spin parameter & N/A \\ \hline 
   {\tt HALO[PX,PY,PZ]} & Comoving 3D halo position & $\hmpc$ \\ \hline 
   {\tt HALO[VX,VY,VZ]} & Physical 3D peculiar halo velocity & $s^{-1}$km \\ \hline
   {\tt HOST\_HALOID} & \texttt{HALOID} of the host halo for subhalos, set to \texttt{HALOID} of the halo for central halos & N/A \\ \hline 
   {\tt XOFF} & Comoving 3D offset of density peak from average particle position & $\hkpc$ \\ \hline 
   {\tt VOFF} & Physical 3D offset of density peak from average particle velocity & $s^{-1}$km \\ \hline 
   {\tt B\_TO\_A} & Ratio of second to first largest halo ellipsoid axis & N/A \\ \hline 
   {\tt C\_TO\_A} & Ratio of third to first largest halo ellipsoid axis & N/A \\ \hline    
   {\tt A[X,Y,Z]} & Direction of the major axis of the halo ellipsoid & N/A \\ \hline \hline  
  \end{tabular}
  \caption{Halo properties}
\end{table}

\section{$N$-body Simulation Methodology}
\label{app:nbody}

\subsection{Lightcone Simulations}

To model large photometric surveys it is necessary to create a dark matter
particle distribution built on a lightcone. We create the lightcones on the fly
as the simulation runs.  Here, every time a particle is moved during the drift
step of the leap-frog integrator, we check to see if it crosses the light cone
surface $r(a)$ --- the comoving distance from the origin of the light cone, $r$,
at the time of the simulation scale factor, $a$.  If a particle on the previous
time step is inside the light cone surface of the previous time step and then is
outside the light cone surface of the current time step, then it has crossed the
light cone surface.  Specifically, for particle $i$ with comoving distances from
the light cone origin on the previous time step $n-1$, $r_{n-1}^{(i)}$ and
similarly for the current time step $n$, $r_{n}^{(i)}$, we check that

\begin{eqnarray}
  r_{n-1}^{(i)} < r(a_{n-1}) &\ \ \ \ \mathrm{and}\ \ \ \  & r_{n}^{(i)} \geq r(a_{n})\nonumber
\end{eqnarray}
where the scale factor of time step $n$ is $a_{n}$.  If a
particle satisfies the light cone crossing condition, we use an interpolation to find the intersection of the
particle's trajectory with the light cone surface.  For each particle,
consider its positions on two consecutive time steps $x_{n}$ and
$x_{n+1}$.  As in \citet{evrard_etal:02}, we define an interpolation
parameter $\alpha$ such that

\begin{displaymath}
x_{\alpha} = x_{n} + \alpha\, v_{n} \, dt\ ,
\end{displaymath}
where $x_{n+1} = x_{n} + v_{n}dt$, $dt$ is the time step, and $v_{n}$ is the
velocity at time step $n$. L-Gadget2 uses a leap-frog integrator and thus this
the velocity is formally displaced by half a time step behind the position. We
label it at the same time step for simplicity.  The light-cone crossing tests
and interpolation are done during the drift step of the leap-frog integrator, so
that as $\alpha$ ranges from zero to one, it traces exactly the particle's
trajectory.

When the particle crosses the light cone surface, it satisfies the following
condition

\begin{displaymath}
|x_{\alpha}|^{2} = r^{2}(t_{n}+\alpha \, dt) \, ,
\end{displaymath}
with $t_{n}$ being the time at time step $n$.  To define the interpolation, at
each time step we compute once an approximation to the function
$r^{2}(t_{n}+dt\alpha)$ defined as

\begin{displaymath}
r^{2}(t_{n}+\alpha \, dt) \approx a\alpha^{2} + b\alpha + c\ .
\end{displaymath}

In our implementation, this approximation is computed through
a least-squares fit of tabulated squared comoving distances which span the range
of the time step $dt$.  We enforce the conditions that at $\alpha=0$ and
$\alpha=1$ the fit produce the correct comoving distances. Thus there is only
one free parameter in the fit for which the least-squares fit is a simple
average, making this procedure extremely efficient. Because the comoving
distance as a function of $\alpha$ is quite smooth, the slight increase in
expense incurred by computing and using the expansion above is offset
by a substantial increase in the accuracy of the interpolation.

With this approximation in hand, the intersection of the particle with the light
cone surface can be computed quickly by solving a quadratic equation

\begin{displaymath}
(dt^{2}|v_{n}|^{2}-a)\alpha^{2} + (2dt\,v_{n}\cdot x_{n}-b)\alpha + |x_{n}|^{2} - c = 0
\end{displaymath}
and selecting the appropriate root.  To compute the scale factor at the
particle's intersection with the light cone surface, we use a tabulated set of
$\{t_{i},a_{i}\}$ values to find $a_{\alpha}$ given $t_{\alpha}=t_{n}+\alpha \, dt$.
We can use this scale factor to test the light cone intersection criterion,
$|x_{\alpha}|^{2}=r^{2}(a_{\alpha})$. We find that this condition is satisfied
to $\lesssim1$ $h^{-1}$kpc.

During the light cone construction, we consider a single fiducial observer at
$(0,0,0)$ in the domain of the simulation cube with length $L$. We create an entire
$4\pi$ steradians of light cone coverage using the periodicity of the
simulation volume to translate the particles into each of the eight cubes which
intersect the (0,0,0) point in the lattice of simulation cubes.  We test for
light cone crossing in each of these eight cubes to generate eight octants which
cover the entire sky.  In practice an extra layer of cubes beyond this fiducial
eight is used to catch extra particles around the corners of the fiducial eight
cubes.  This extra layer is discarded once the light cone surface has moved
sufficiently into the interior of the volume of the eight fiducial cubes.
Particles can be output twice into the same octant at different scale factors in
this scheme. Our procedure ensures that the large-scale structure is continuous
between the different octants of the light cone.  Additionally, by choosing
carefully which particles to save, we can generate pencil-beam light cones of
different geometries if desired.

The procedure described above generates a full-sky light cone out to comoving
distance $L$ for a simulation with side length $L$.  This full sky light cone
has repetitions out to comoving distance $L$, but encloses a unique simulation
volume out to a comoving distance of $L/2$, and any given octant is unique out
to a distance of $L$.  The full sky light cone is used for the weak-lensing ray
tracing calculations as described below to achieve the proper boundary
conditions.

\subsection{High-Resolution Tuning Simulation}

We run a single higher-resolution simulation used to tune
the \textsc{Addgals} galaxy assignment algorithm.  For this, we require a simulation with
sufficient resolution to use the subhalo abundance matching technique (SHAM; see
e.g. \citealt{Conroy2006,Reddick12}) to model the galaxy distribution down to roughly $M_r= -19$.  To do this we use a simulation box of size $400~\hinv$Mpc with $2048^3$
particles. At this resolution, the SHAM catalog is not strictly complete down to
-19, as subhalos near the cores of massive hosts are stripped below the
simulation resolution (see \citealt{Reddick12} for a detailed discussion).
However, comparisons with SDSS data show that the resolution is sufficient to
model the observed 2-point function to $r_{p} \sim 300 \hkpc$ to within current observational errors down to  $M_r = -19$, and moderately well below this limit \cite{wechsler_etal:19}. A lightcone output is not necessary for this computational volume, but we require merger trees to construct the abundance matching catalog. For this, we save 100 simulation snapshots logarithmically spaced from z = 12 to z = 0, which allows construction of accurate merger trees.

\subsection{Halo Finding}
\label{sec:halofinding}

Halo finding is done with the publicly available adaptive phase-space halo
finder {\sc  Rockstar}\footnote{https://bitbucket.org/gfcstanford/rockstar}
\citep{behroozi_etal:13a}.  {\sc Rockstar} is very efficient and accurate (see for example, the halo finder comparison in \citealt{Knebe11}).  It is particularly robust in galaxy mergers, important for the massive end of the halo mass function, and in tracking substructures, important for the abundance matching procedure applied to the tuning simulation.  The total number of halos with more than 100 particles found in the lightcone volume used for each simulation is given in Table
\ref{table:simulations}. We have chosen to use $\mathrm{M}_\textrm{vir}$ strict spherical overdensity (SO) masses \citep{bryan_norman:98}; additional halo mass definitions are output by {\sc Rockstar} using these centers.  {\sc Rockstar} also outputs several other halo properties, including other halo mass measurements, concentration, shape, and angular momentum \citep[see][for details]{behroozi_etal:13a}.

Comparisons between our simulations and standard halo mass function
\citep{McClintock2018} and halo bias \citep{Tinker2010} models are shown in
Fig.~\ref{fig:halo_comp}. In both cases measurements are averaged over 3 sets
of $10,313$ square degree simulations and the error bars plotted are the error
on the mean estimated via jackknife. The mass functions in L1 (measured for
$0.0<z\leq 0.34$) and L2 (measured for $0.34<z\leq 0.9$) agree very well with
\citet{McClintock2018}, with discrepancies at low mass likely due to differences
in halo finding, as the simulations used in \citet{McClintock2018} identified
halos using strict spherical overdensity estimates around centers defined using
$M_{200b}$ rather than re-measuring $M_{200b}$ around centers defined when finding halos using $M_{\textrm{vir}}$ as the halo definition in this work. The mass functions in
L3 deviate from the emulator prediction in a mass dependent way that is likely
attributable to the low mass resolution in these simulations. The right
panel of Fig.~\ref{fig:halo_comp} shows halo bias measurements for a bin of
mean mass equal to $4\times 10^{13} \hmsun$ in the L1 ($z=0.26$), L2
($z=0.54$) and L3 ($z=1.34$) simulations compared to the \citet{Tinker2010}
bias model. These measurements as well as measurements we performed at higher
masses and different redshifts agree to within the quoted $6\%$ error on the
model.

\subsection{Merger Tree}

For the highest resolution ``tuning simulation'', we track the formation of
halos using 100 saved snapshots between $z = 12.3$ and $z = 0$, equally spaced
in $\Delta \ln a$.  The gravitationally-consistent merger tree
algorithm\footnote{https://bitbucket.org/gfcstanford/consistent-trees} described in
\cite{behroozi_etal:13b} is applied to track halos.  This algorithm explicitly
checks for consistency in the gravitational evolution of dark matter halos
between time steps, and leads to very robust tracking.  Details of the
implementation and its robustness can be found in \cite{behroozi_etal:13b}.
Using the resulting merger trees, we are able to track ${\rm v}_{\rm max}$ and
${\rm v}_{\rm vir}$ at $M_{\rm peak}$ for each identified subhalo. The quantity

\begin{equation}
v_{\alpha}=\left (\frac{v_{\textrm{max}}}{v_{\textrm{vir}}}\right)^{\alpha}
\end{equation}
with $\alpha=0.68$ is used to assign galaxies to dark matter halos, using an
abundance matching algorithm described by \cite{Lehmann2017} with 0.17 dex
scatter in absolute magnitude.

\section{Weak Gravitational Lensing Implementation}
\label{app:weaklens}

We calculate the shear and magnification applied to the galaxy by weak
gravitational lensing.  To do this calculation, we use the multiple-plane ray
tracing code \textsc{CALCLENS} described in \cite{becker:12}. With a
multiple-plane ray tracing code, we can correctly find the lensed images of the
galaxies (including multiply imaged objects) and we naturally include
higher-order corrections to the Born approximation in the shear and convergence
fields \citep[e.g.,][]{hilbert2009}.  Multiple-plane ray tracing codes use the
Limber approximation to relate the surface mass density in radial shells along
the line-of-sight of the light cone to the lensing potential in each shell
through a two-dimensional Poisson equation \citep[e.g.,][]{jain2000}. Once this
two-dimensional Poisson equation is solved, the derivatives of the lensing
potential are then used to propagate the ray locations and their inverse
magnification matrices shell to shell from the observer to the furthest edge of
the light cone.

The multiple-plane ray tracing code \textsc{CALCLENS} of \cite{becker:12} tracks
both the ray positions and the inverse magnification matrix at each ray,
correctly accounts for the sky curvature in the Limber approximation, uses
HEALPix\footnote{http://healpix.jpl.nasa.gov/} \citep{gorski2005} for the ray
locations to achieve uniform resolution over the sphere, and finds the galaxy
images using a grid search algorithm \citep[e.g.,][]{schneider1992,hilbert2009,Fosalba2008}
implemented on the sphere. A grid search algorithm is capable of correctly
finding multiple images of the same source. However, the lens causing the
multiple images must be properly resolved in order for these multiple images to
be correct. Thus given the resolution of the simulations used in this work, in
practice very few multiple images are found for a given source galaxy catalog
and those that are found are not expected to be computed accurately. Also, as
implemented currently, only the average position of the multiple images is
computed, so that strong lensing features like arcs are not captured in the
catalogs. While CALCLENS has the capability to use a combination SHT+Multigrid
Poisson solver for the sphere, for this work we choose to use just the SHT
version as it is slightly more accurate. Once the inverse magnification matrix
and lensed position of each source galaxy is computed, they are then used to
lens the source galaxy catalog with its shapes and sizes and to apply the
magnification to the magnitudes of each galaxy as described above.

The code is written in \verb+C+, uses common software packages
(GSL\footnote{http://www.gnu.org/software/gsl},
FFTW3\footnote{http://www.fftw.org},
FITS\footnote{http://fits.gsfc.nasa.gov/iaufwg/iaufwg.html},
HDF5\footnote{http://www.hdfgroup.org/HDF5}) and is MPI-parallel so that it is
quite portable and efficient.  This code is publicly available for other
researchers.\footnote{https://github.com/beckermr/calclens}

For these simulations, we perform the ray tracing using an $n_{side}=8192$ grid
resulting in an effective pixel size of $\theta_{pix}=0.46'$.
The right panel of Fig.~\ref{fig:nbody_comp} shows measurements of $\xi_{+/-}$ without shape
noise in three redshift bins averaged over the 18 Y1 footprints presented in
this work compared to an analytic prediction for $\xi_{+/-}$ from
\textsc{halofit} \citep{takahashi2012}. The bottom panel shows the fractional
deviation of the simulations from the analytical predictions, with the grayed
out regions demarcating scales below the pixel size used to do the raytracing.
This resolution is sufficient for modeling $\xi_{+/-}$ at the scales used in the
DES Y1 analysis, which are shown by the blue and green vertical lines. The
dashed lines in the bottom panel show

\begin{equation*}
\Delta \xi_{+/-}(\theta) = \frac{\xi_{+/-,\ell<\ell_{max}} - \xi_{+/-,\textsc{halofit}}}{\xi_{+/-,\textsc{halofit}}}
\end{equation*}

where

\begin{align}
\xi_{+/-,\ell<\ell_{max}}(\theta)
=&\int \frac{d\ell \,\ell}{2\pi}\,\frac{J_{0/4}(\ell\theta)}{1 + \exp \left [\textrm{ln}(\ell) - \textrm{ln}(\ell_{max})\right ]}  \nonumber \\
& \int d\chi  \frac{q_\kappa^i(\chi)q_\kappa^j(\chi)}{\chi^2}P_{\textsc{HALOFIT}}\left(\frac{\
l+1/2}{\chi},z(\chi)\right)
\end{align}

with
\begin{align}
q_\kappa(\chi) = \frac{3 H_0^2 \Omega_m }{2 \mathrm{c}^2}\frac{\chi}{a(\chi)}\int_\chi^{\chi_{\mathrm{h}}} \
d \chi' \frac{n_{\kappa} (z(\chi')) dz/d\chi'}{\bar{n}_{\kappa}} \frac{\chi'-\chi}{\chi'}
\end{align}

i.e. $\xi_{+/-,\ell<\ell_{max}}(\theta)$ is the \textsc{halofit} prediction but
with the integral over $\ell$ truncated by an exponential function above a
characteristic scale $\ell_{max} = \frac{\pi}{\theta_{\textrm{pix}}}$. This
truncation is an approximation to the effect of resolution in the ray tracing
algorithm, since the finite pixel size used to perform the raytracing
calculations will manifest itself as an effect with a constant angular scale.
The fact that the angular scale at which resolution effects begin diminishes as
a function of redshift, instead of remaining constant, indicates that the
resolution effects are sourced by effects in the underlying density field and
not the raytracing itself. At low redshift, the fact that the deviation of the
simulations from convergence matches the prediction using the truncation in
$\ell$ as described above is likely a red herring produced by the fact that the
physical scales that suffer from resolution effects in the lightcones translate
to roughly the same angular scale at these redshifts as $\theta_{pix}$.

Finally, once we have assigned magnitudes, shapes, and sizes to the galaxies,
we lens the shapes using the relations

\begin{equation}
  \label{eq:6}
    \varepsilon =
  \begin{cases}
    \frac{\varepsilon^{\mathrm{(s)}} + g}{1 + g^*\varepsilon}  & |g|
   \le 1 \; , \\
   \frac{1 -
   g\varepsilon^{*\mathrm{(s)}}}{\varepsilon^{*\mathrm{(s)}} +
  g^*} & |g| > 1 \; ,
\end{cases}
\end{equation}

\citep{seitz_schneider:97} where the asterisk denotes complex
conjugation, $g$ is the complex reduced shear computed from the
ray-tracing and the superscript $(s)$ here and in the following
denotes intrinsic source properties. The sizes and magnitudes are
changed by gravitational lensing according to

\begin{equation}
  \label{eq:7}
  r = \sqrt{|\mu|} r^{\mathrm{(s)}}\;,
\end{equation}

and

\begin{equation}
  \label{eq:8}
  m = m^{\mathrm{(s)}} - 2.5 \log(|\mu|)\;,
\end{equation}

where $\mu$ is the lensing magnification. Lensing deflections are also added to the true angular coordinates of each galaxy:

\begin{align}
RA^{\prime} =  RA + \delta_{RA} \\
DEC^{\prime} = DEC + \delta_{DEC}
\end{align}
where $RA^{\prime}$ and $DEC^{\prime}$ are the lensed coordinates and RA and DEC are the un-lensed coordinates.

\section{Dark Energy Simulations with Second-order Lagrangian
         Perturbation Theory Initial Conditions}\label{app:de2lptic}

Although the catalog presented here uses a $\Lambda$CDM cosmology, In
this work, we use second-order Lagrangian perturbation theory (2LPT)
initial conditions (ICs) for our numerical simulations (see
Section~\ref{app:nbody}). Our simulations implement changes to the
background expansion rate (which subsequently changes the rate of
growth of structure) at late times only and neglect any effects of
dark energy perturbations.  As pointed out by \citet{alimi2010}, this
model is phenomenological rather than based on a single underlying
theory.  Previous works have used a rescaling of \lcdm\ ICs
\citep[e.g.,][]{dolag2004} or exact integrations of the first-order,
linear growth equations for DE models with ZA ICs
\citep[e.g.,][]{alimi2010}. Here we present the ordinary differential
equations (ODEs) for the first- and second-order growth factors along
with prescriptions to integrate them for general DE models which
change the background expansion rate only. These are the numerical
coefficients needed for implementing 2LPT ICs with general DE models.

We start with equations E.18 and E.31 of Appendix E of \citet{jeong2010}. These
equations, when transformed to be functions of time $t$ instead of the conformal
variable $d\tau=dt/a$, are

\begin{eqnarray}
\ddot{D}_{1} + 2H(a)\dot{D}_{1} - 4\pi G\rho_{m}(a)D_{1}& = & 0\\
\ddot{D}_{2} + 2H(a)\dot{D}_{2} - 4\pi G\rho_{m}(a)D_{2}& = & -4\pi G\rho_{m}(a)D_{1}^{2}\ ,
\end{eqnarray}
where $D_{1,2}$ are the first- and second-order growth factors, $H(a)$ is the
Hubble function, $a$ is the scale factor normalized to unity today, and
$\rho_{m}(a)\propto a^{3}$ is the mean matter density. The dots denote
derivatives with respect to time $t$. The most convent form for numerical work
is to apply an additional transformation to make them functions of the scale
factor $a$. Additionally, it is convenient to express the mean matter density in
units of the critical density so that $4\pi
G\rho_{m}(a)=\frac{3}{2}H(a)^{2}\Omega_{m}(a)$. The final results are for
$D_{1}$,

\begin{eqnarray}
\frac{dD_{1}}{da} & = &D_{1}'\\
\frac{dD_{1}'}{da} & = &\frac{3}{2}\frac{\Omega_{m}(a)}{a^{2}}D_{1} - \left[\frac{3}{a} + \frac{d(H^{2}(a))}{da}\frac{1}{2H^{2}(a)}\right]D_{1}'\ ,
\end{eqnarray}

and for $D_{2}$,

\begin{eqnarray}
\frac{dD_{2}}{da} & = &D_{2}'\\
\frac{dD_{2}'}{da} & = &\frac{3}{2}\frac{\Omega_{m}(a)}{a^{2}}D_{2} - \left[\frac{3}{a} + \frac{d(H^{2}(a))}{da}\frac{1}{2H^{2}(a)}\right]D_{2}' - \frac{3}{2}\frac{\Omega_{m}(a)}{a^{2}}D_{1}^{2}\ .
\end{eqnarray}

Here the primes denote the derivative of the growth factors. The second-order
ODEs have been split into two first-order ODEs so that they can be integrated
numerically. We use a high-order method from the publicly GNU Scientific
Library\footnote{\url{https://www.gnu.org/software/gsl/}} to integrate these
systems of ODEs.

To complete the integrations, one needs ICs for both $D_{1,2}$ and their
derivatives. To do this, we follow \citet{WMAP5} and specify them in the matter
dominated era where exact solutions are know and the simulations are started.
These solutions are \citep[e.g.,][]{bouchet1995}

\begin{eqnarray}
D_{1}(a_{i}) &=&a_{i}\\
D_{1}'(a_{i}) &=&1\\
D_{2}(a_{i}) &=&-\frac{3}{7}a^{2}_{i}\\
D_{2}'(a_{i}) &=&-\frac{6}{7}a_{i}\ .
\end{eqnarray}

With these ICs, one obtains as the solution the strongest growing mode out of
all of the homogenous and, for the second-order growth factor $D_{2}$,
homogeneous and particular solutions to the above equations. These modes are the
appropriate solutions for initializing cosmological simulations.

Finally, one typically works with growth factors normalized to unity at the
current epoch. For the linear order growth factor $D_{1}$, one simply normalizes
via $\widetilde{D}_{1}(a)\rightarrow D_{1}(a)/D_{1}(a=1)$. Here
$\widetilde{D}_{1,2}$ denotes the normalized growth factors. For the
second-order growth factor, this normalization corresponds to
$\widetilde{D}_{2}(a)\rightarrow D_{2}(a)/D_{1}^{2}(a=1)$. To see this, remember
that $\widetilde{D}_{2}(a)$ will be multiplied by a quantity which is
$\propto\delta(k)^{2}$. Thus if $\delta(k)$ is normalized such that
$P(k,a=1)=\frac{1}{(2\pi)^{3}\delta(k+k')}\langle\delta(k)\delta(k')\rangle$ is
the linear power spectrum today, then one must divide by $D_{1}^{2}(a=1)$ to
obtain the proper normalization.

\section{High-redshift ADDGALS Modifications}
\label{app:gals}

We use the \addgals\ algorithm, as described in \cite{wechsler_etal:19}, to
create a galaxy catalog. Here we summarize the extensions to this algorithm that 
are required to model higher redshift observables than discussed 
in \cite{wechsler_etal:19}. \addgals\ is an
empirical method for generating a catalog of galaxies to generate the
distribution of galaxies and a minimal set of their photometric and
spectroscopic properties. The algorithm matches the clustering properties
predicted by SHAM, but without the use of high resolution halo merger histories.
This is accomplished through a model fitting process to a SHAM catalog on our
smaller, high resolution \textsc{T1} simulation.  This model is then used
to populate our lower resolution lightcone simulations, \textsc{L1, L2, L3},
with galaxies with rest frame $r$-band absolute magnitudes, described in Section \ref{sec:galaxies}. A second step of the algorithm assigns SEDs to each galaxy 
allowing for the calculation of observed frame magnitudes as described in \ref{sec:colors}.

\subsection{Galaxy Luminosity Function}
\label{sec:lf}

The main input that \addgals\ uses to generate galaxy populations is a rest
frame $r$-band luminosity function as a function of redshift, $\phi(M_{r}, z)$.
There are large systematic discrepancies between many luminosity function
measurements due to differences in measuring photometry \citep{Bernardi2012},
correcting for incompleteness in flux \citep{Blanton2005} and sample variance of
the small fields used at higher redshift \citep{Loveday2015}. Due to these
issues, we focus on matching the cumulative number of observed counts as a
function of apparent $griz$ magnitudes, $\vec{n}(>\vec{m})$ in DES Y1, a
directly observable quantity. This quantity has limited constraining power on
the luminosity function as a function of redshift, the quantity that \addgals\
requires as an input. Because of this, instead of constraining the full shape of
$\phi(M_{r},z)$  using $\vec{n}(>\vec{m})$, we start by taking the shape of the
luminosity function as measured to high precision at low redshift from the work
using the method described in \cite{Reddick12} based on the SDSS spectroscopic sample and use the measurements of $\phi_{*}(z)$ from AGES \citep{Cool2012} to
account for redshift evolution. We then allow for additional freedom in
$M_{*}(z)$, which we constrain using $\vec{n}(>\vec{m})$.

In particular, the functional form of the luminosity function that we use is
a modified double-Schechter function with a Gaussian tail, as given by

\begin{align}
\Phi(M,z) &= 0.4 \ln(10) e^{-10^{-0.4(M-M_*(z))}}
\left(\phi_1(z) 10^{-0.4(M-M_*(z))(\alpha_1 + 1)} + \phi_2(z) 10^{-0.4(M-M_*(z))(\alpha_2 + 1)} \right) \\
& + \nonumber {\phi_3(z) \over \sqrt{2\pi \sigma_{hi}^2}} e^{{-(M-M_{hi}(z))^2 \over 2 \sigma_{hi}^2}}.
\label{eq:dsg}
\end{align}

where
\begin{align}
M_{*/hi}(z) = M_{*/hi,0} + Q\left (\frac{1}{1+z} - \frac{1}{1.1}\right)
\end{align}

and

\begin{align}
\phi_{i}(z) = \phi_{i,0} + Pz
\end{align}

We first produce a catalog of galaxies with the luminosity function model as
described above with the values for $Q$ and $P$ taken from \citet{Cool2012}, where the same values for $Q$ and $P$ are used for $M_{*}$ and $M_{hi}$ and all the $\phi_{i}$ respectively. The double Schechter function is motivated by evidence that the faint end of the luminosity function deviates from a single power law \citep[e.g.]{blanton_etal:03}. The Gaussian is intended to model observed deviations from the exponential fall off prescribed by a Schechter function at the bright end \citep[e.g.]{Reddick12, Bernardi2012}.

We then parameterize the discrepancy between our luminosity function model and the
luminosity function of DES galaxies as a simple shift in $M_*$ as a function of
z

\begin{equation}
\Delta M_*(z) =  \frac{\Delta Q}{1.1}\frac{0.1 - z}{1+z}
\end{equation}

and fit for $\Delta Q$ by maximizing the log likelihood given by

\begin{equation}
\log \mathcal{L} \propto \left[\vec{n}^{DES} - \vec{n}(\Delta Q)\right] \Sigma^{-1} \left[\vec{n}^{DES} -\vec{n}(\Delta Q)\right]^{T}
\end{equation}

where $\vec{n}^{DES}$ is the cumulative number of galaxies as a function of
magnitude per square degree measured in the DES overlap with COSMOS (which is
approximately 1 magnitude deeper than the DES Y1 wide field). $\vec{n}(\Delta
Q)$ is the cumulative number of galaxies per square degree in Buzzard when
applying a shift to each galaxy's $griz$ magnitudes given by $\Delta M_*(z)$.
$\Sigma$ is approximated as diagonal and estimated as the Poisson error on
$\vec{n}^{DES}$ and $\vec{n}(\Delta Q)$ added in quadrature. Magnitudes ranging from $\{18.6, 17.8,17.4, 17.1\}$ down to magnitudes $\{24.38, 24.25, 
23.71, 23.26\}$ in $g$, $r$, $i$, and $z$-band respectively are used in the fits. The bright end of this cut is imposed to mitigate the large amount of sample variance
in the $\sim 1.5$ square degree COSMOS field. At the faint end, there should
be $>10\%$ contributions from galaxies with $z>2.3$, which are not able to
 be placed in our simulations.

\subsection{Galaxy SEDs and Multi-band Photometry}
\label{sec:colors}

We map galaxy SEDs from the data onto our simulated galaxies by using the relation between $M_r$, projected galaxy density, and SED as measured from SDSS. Using a set of spectroscopic galaxies comprising 570,000 objects
from the SDSS VAGC \citep{blanton05} with $0.005<z<0.2$, we measure $\Sigma_5$,
the projected distance to the $5^{th}$ nearest galaxy brighter than
$M_{r}=-19.75$ in redshift slices with $\Delta z=0.02$. Similar quantities have been shown to correlate strongly with star formation rate \citep{Cooper06}. In
bins of $M_{r}$ and $z$ we then rank each galaxy by this density, yielding the rank
$\mathcal{R}_{\Sigma_5}$. By using $\mathcal{R}_{\Sigma_5}$ instead of
$\Sigma_5$, we avoid issues related to evolving number densities that we would
otherwise encounter due to our use of a magnitude limited, as opposed to volume
limited, sample for our training set.

Identical measurements are made for our simulated galaxies and for each
simulated galaxy, $g_i$, with absolute magnitude $M_{r,i}$, redshift $z_i$ and
density rank $\mathcal{R}_{\Sigma_5, i}$ we identify a set of possible matches
in the data, $\{g_{j,\textrm{SDSS}}\}$, by selecting galaxies in the data in the
same $M_{r}-\mathcal{R}_{\Sigma_5}$ bin. If we were modeling the same redshift
range as our training set, we could sample uniformly from this set to draw an
SED for $g_i$. This is not the case in this work, as we wish to model a much
larger redshift range than the training set we are using. Thus we must account
for redshift evolution of the $M_{r}-\mathcal{R}_{\Sigma_5}-\textrm{SED}$
relationship. We do this by assuming that our training set spans the set of
SEDs that may appear in DES, but allow for the possibility that the 
distribution of these SEDs evolves as a function of redshift and absolute magnitude.
To model this evolution, we define the quantity $\mathcal{W}_{\textrm{red}}(M_{r},z)$, as ratio of the red fraction of galaxies at redshift $z$ relative to the low
redshift red fraction (at $z < 0.2$). A galaxy is deemed to be on the red
sequence if it satisfies

\begin{equation}
^{0.1}(g-r) > 0.15 - 0.03^{0.1}M_r.
\end{equation}

We then define

\begin{align}
\mathcal{W}_{\textrm{red}}(M_r,z) &= \frac{P(\textrm{red} |z, M_{r}; \textrm{PRIMUS})}{P(\textrm{red} | 0.0<z<0.2, M_r; \textrm{PRIMUS})},
\label{eq:fql_weight}
\end{align}

i.e. the ratio of the red fraction as a function of $r$-band absolute magnitude in
PRIMUS \citep{primus}, to that found at low redshift in PRIMUS. We then choose a
galaxy from the set of matches with SDSS, $\{g_{j,\textrm{SDSS}}\}$, choosing a
red galaxy with probability given by

\begin{equation}
P(\textrm{red}|M_r,z,\mathcal{R}_{\Sigma_5})=P(\textrm{red}|M_r,z,\mathcal{R}_{\Sigma_5};\textrm{SDSS})\mathcal{W}_{red}(M_r,z)
\end{equation}

and assign its SED to $g_i$ using \textsc{kcorrect} coefficients.
$\textsc{kcorrect}$ is a SED template fitting method described in
\citet{blanton_etal:03kcorr} that represents SEDs as a linear combination of 5
templates which were tuned to match SDSS $griz$ photometry. Using this SED
representation, we are then able to generate multiband photometry for each
galaxy.

\subsection{Adding Realistic Galaxy Shapes and Sizes}
\label{sec:lens}

Gravitational lensing changes the shape, size, and position of galaxy images. To
include the effects on size and shape in our simulations we first need a model
for the intrinsic distributions of these quantities before the effects of
lensing are applied to them. We are primarily interested in modeling weak
gravitational lensing, and we limit the shape model to ellipticities and neglect
higher order terms describing more complex galaxy shapes.

In real imaging data a distinction has to be made between an object's
ellipticity and the shear estimated from its observed ellipticity. Shear
estimation methods call this e.g. the shear polarizability
\defcitealias{kaiser_etal:95}{KSB}\citep[][KSB]{kaiser_etal:95}. For purely
catalog based simulations such a distinction is not necessary and we base our
model for the intrinsic ellipticity of objects on the shear estimators reported
by the \citet{Zuntz2017} implementation of the \textsc{metacalibration} algorithm. We also wish to model galaxy angular sizes, by re-sampling the SExtractor \citep{bertin_arnouts:96} measurements of \texttt{FLUX\_RADIUS} in the same DES Y1 data set. The distribution function of these shear estimators,
which we use as intrinsic ellipticity distribution, and sizes, is described by an
Gaussian mixture model. In particular we fit a Gaussian mixture with 20 components to the joint distribution of absolute ellipticity, \texttt{FLUX\_RADIUS}, and $i$-band magnitude in the DES Y1 \textsc{metacalibration} sample, deconvolving magnitude errors using the method described in \citet{Bovy2009}.

\begin{equation}
  \label{eq:1}
  p(|\varepsilon|, r, m_{i}) = \sum_{i=0}^{19} \alpha_{i}\mathcal{N}(\mu_{i}, \Sigma_{i})
\end{equation}

Here $|\varepsilon|$ is the absolute value of the ellipticity, $r$ is the \texttt{FLUX\_RADIUS}, $\mu_{i}$ and $\Sigma_{i}$ are each Gaussian components' mean and covariance matrix, and $\alpha_{i}$ are the component weights, which sum to unity. Ellipticities and sizes are then drawn for each galaxy by conditioning \ref{eq:1} on the galaxies $i$-band magnitude and sampling from the resulting conditional distribution.

\subsection{Photometric errors}
\label{sec:photoerr}

Once the catalogs have been lensed and are rotated into the DES Y1 footprint, we
apply a straightforward model to include photometric errors. Photometric errors
provide a significant source of contamination, particularly for apparently faint
galaxies.  These errors will cause both objects above the detection threshold to
scatter out of our detection limits, as well as causing many more dim objects to
scatter in.  Modeling these errors appropriately can thus be important for a
number of analyses.

To match the DES Y1 magnitude and magnitude error distribution we use the map of
10-$\sigma$ limiting magnitudes determined from the DES Y1 Multi-Object Fitting
(MOF) photometry in $griz$ as well as the effective exposure time
($t_{\mathrm{eff}}$) of the survey in each band. In this way we realistically
incorporate depth variations from the Y1 footprint in our simulations,
neglecting the dependence of detection efficiency on galaxy size. Our model for
photometric errors using these ingredients is as follows.

We use a straightforward method of calculating the Poisson noise for the flux of
a simulated galaxy plus the sky noise in a particular band.  Here, the total
signal from these two sources (galaxy and sky) are given by the relation:

\begin{equation}
  \begin{split}
    S_{\mathrm{gal}} &= 10^{-0.4(m_\mathrm{gal} - ZP)}\times t_{\mathrm{eff}} \label{eq:flux_gal}\\
    S_{\mathrm{sky}} &= f_{\mathrm{sky}} \times t_{\mathrm{eff}},
  \end{split}
\end{equation}

where $m_{\mathrm{gal}}$ is the magnitude of a galaxy and $f_{\mathrm{sky}}$ is the sky noise (in a particular band), and $t_{\mathrm{eff}}$ is the effective exposure time in nominal observing conditions as defined in \citet{y1gold}. In all cases we set the zero-point $ZP = 22.5$, and all fluxes
in the data tables are converted to nanomaggies such that:

\begin{equation}
  m = 22.5 - 2.5\log_{10}f_{\mathrm{nmgy}}.
\end{equation}

Finally, we note that the sky noise parameter, $f_{\mathrm{sky}}$, can be
estimated from the $10\sigma$ limiting magnitude $m_{\mathrm{lim}}$ and the
associated $f_{\mathrm{lim}}$ :

\begin{equation}
  f_{\mathrm{sky}} = \frac{f_{\mathrm{lim},1}^2 \times t_{\mathrm{eff}}}{100} -
  f_{\mathrm{lim},1},
\end{equation}

where $f_{\mathrm{lim},1}$ is the 1~second flux at the limiting magnitude given
by Eqn.~\ref{eq:flux_gal}. Given the galaxy flux and sky flux, the typical noise
associated with each galaxy will be given by a random draw from a distribution
of width $\sigma_{\mathrm{flux}} = \sqrt{S_{\mathrm{gal}} + S_{\mathrm{sky}}}$.
Then, for each galaxy, we convert the total observed flux and error to
nanomaggies, such that $f_{\mathrm{nmgy}} =
S_{\mathrm{gal,obs}}/t_{\mathrm{eff}}$. Finally, we calculate the magnitude and
error:

\begin{equation}
  \begin{split}
    m_{\mathrm{obs}} &= 22.5 - 2.5\log_{10}(f_{\mathrm{nmgy}}) \\
    m_{\mathrm{err,obs}} &= \frac{2.5}{\ln(10)}\frac{f_{\mathrm{err,nmgy}}}{f_{\mathrm{nmgy}}}
  \end{split}
\end{equation}

Note that, as is possible with observational data, this results in a number of
dim galaxies having negative fluxes in one or more bands in cases with large
photometric errors.  These are set to $99$ in the magnitude table.

\section{Differences from previous catalog versions}
Previous versions of these catalogs contained a large bias in the mean redshift of the fourth source bin that was significantly ruled out by the data \citep{MacCrann17}. The selection used for those source catalogs was
\begin{enumerate}
\item Mask all regions of the footprint where limiting magnitudes and PSF sizes cannot be estimated.
\item $m_r < -2.5 \log_{10}(1.5) + m_{r, lim}$
\item $\sqrt{r_{gal}^{2} + (0.5~ r_{PSF})^{2}} > 0.75~r_{PSF}$
\item $m_r < 22.07 + 1.08~z$
\end{enumerate}

The source of this bias is our use templates constrained by SDSS when incorporating colors in our simulations. These templates are constrained by SDSS $griz$ photometry at $z\le0.2$. Because of this, the rest frame UV portions of the templates are relatively unconstrained, and it is this portion of our spectra which appear in the DES band-passes at $z>\sim 1.5$, leading to little evolution in color
of our galaxies above this redshift. As such, BPZ cannot distinguish between
galaxies at $z=1.5$ and higher redshifts, leading to many higher redshift 
galaxies contaminating the highest redshift source redshift bin 
($0.9<z_{BPZ}\le 1.3$). The previous version of these catalogs, \textsc{Buzzard v1.6} had significantly more bright high redshift galaxies than the current version, exacerbating the problem with high redshift colors. High redshift SEDs in our
simulations represent one of the major outstanding deficiencies of these simulations, and we are actively pursuing upgrades which will improve this performance. 
Until then, high-redshift galaxy colors in these simulations should be
used with caution.

\bibliographystyle{yahapj}
\bibliography{adstex}

\end{document}